\documentclass[10pt]{iopart}
\usepackage{epsfig,graphicx, setspace}
\usepackage{wrapfig}

\newcommand{\room}{\rule[-0.3cm]{0cm}{0.8cm}}

\newcommand{\vsp}{\vspace*{3mm}}
\newcommand{\be}{\begin{equation}}
\newcommand{\ee}{\end{equation}}
\newcommand{\bd}{\begin{displaymath}}
\newcommand{\ed}{\end{displaymath}}

\newcommand{\extr}{~{\rm extr}}
\newcommand{\bra}{\langle}
\newcommand{\ket}{\rangle}

\newcommand{\order}{{\cal O}}
\newcommand{\minus}{\!-\!}
\newcommand{\plus}{\!+\!}

\newcommand{\bc}{\mbox{\boldmath $c$}}

\newcommand{\bk}{\mbox{\boldmath $k$}}

\newcommand{\bomega}{\mbox{\boldmath $\omega$}}

\newcommand{\bpsi}{\mbox{\boldmath $\psi$}}

\newcommand{\bki}{\vec{k}_i}
\newcommand{\bkj}{\vec{k}_j}

\newcommand{\squishlist}{
 \begin{enumerate}
  { \setlength{\itemsep}{0pt}
     \setlength{\parsep}{3pt}
     \setlength{\topsep}{3pt}
     \setlength{\partopsep}{0pt}
     \setlength{\leftmargin}{1.5em}
     \setlength{\labelwidth}{1em}
     \setlength{\labelsep}{0.5em} } }
\newcommand{\squishend}{
  \end{enumerate}  }
  
\begin{document}
\eqnobysec

\title[Tailored graph ensembles as proxies or null models for real networks]
{Tailored graph ensembles as proxies or null models for real networks II: results on directed graphs}

\author{ES Roberts$^{\dag\ddag}$, ACC Coolen$^{\dag\ddag\S}$, and T Schlitt$^{\star}$} 

\address{
${\dag}~$Department of Mathematics, King's College London, The Strand,
London WC2R 2LS, United Kingdom}
\address{
${\ddag~}$Randall Division of Cell and Molecular Biophysics, King's College London, New
Hunts House, London SE1 1UL, United Kingdom}
\address{$\S~$
London Institute for Mathematical Sciences, 35a South St, Mayfair, London W1K 2XF, United Kingdom}
\address{
$\star~$ Department of Medical and Molecular Genetics, King's College London School of Medicine, 8th floor Guy's Tower, London SE1 9RT, United Kingdom}

\begin{abstract} 
We generate new mathematical tools with which to quantify the macroscopic topological structure of large directed networks. This is achieved via a statistical mechanical analysis of constrained maximum entropy ensembles of directed random graphs
 with prescribed joint distributions for in- and out-degrees and prescribed degree-degree correlation functions.  
We calculate exact and explicit formulae for the leading orders in the system size of the Shannon entropies and complexities of these ensembles, and for
information-theoretic distances.
 The results are applied to data on gene regulation networks.
 \end{abstract}
 
  \pacs{ 87.18.Vf, 89.70.Cf, 89.75.Fb, 64.60.aq\hspace*{\fill} }

\ead{ekaterina.roberts@kcl.ac.uk, ton.coolen@kcl.ac.uk, thomas.schlitt@kcl.ac.uk}

\section{Introduction}

There is a great demand, especially in cellular biology,  for 
 precise mathematical tools with which to quantify topological structure in large observed networks.  
 Such tools can be used to: compare
networks; distinguish between meaningful and random structural features; and, to define and generate 
 tailored random graphs as null models or network proxies. 
In a previous paper   \cite{AnnCooFerFraKle09} it was shown how a specific family of tailored random graph ensembles, with controlled 
degree distributions and controlled degree-degree correlation functions, is well suited for generating such tools. 
 The authors of   \cite{AnnCooFerFraKle09}  applied techniques from statistical mechanics 
 to calculate explicit formulae for 
  the leading orders in the systems size of the Shannon entropy per node  for these tailored graph ensembles, and related quantities such as complexity and information-theoretic distances. 
Subsequent papers were devoted to the numerical generation of  graphs \cite{CooDemAnn09} from the proposed ensemble families and the
 application in cellular biology of the resulting mathematical tools   \cite{FernandesETAL2010}. For an overview see e.g. \cite{CoolenETAL2010}.
The main limitation of \cite{AnnCooFerFraKle09} was that it only dealt with nondirected networks and graphs. In this paper we take the next 
step and develop the corresponding theory for directed ones.   

Extending the methods in \cite{AnnCooFerFraKle09}  to directed networks will enable their application to important new problems especially in cellular biology. Other applications could include the analysis and control of communication and computation networks. For example, to understand the processes driving a cell it is necessary to go beyond studying individual genes; one needs to study their interactions. Information on how genes interact within the cell is commonly represented by a directed graph: the gene regulation network. High-throughput methods have generated a wealth of data on gene regulation. We now need powerful mathematical tools to analyse these data. By focussing on which properties are the most important to the structure of the biological signalling network, we can envisage being able to postulate mechanisms for how the network evolved and came to fulfil its function, and 
build better models for such networks. 
Evaluating the fit of a network model to network data is often seen as a formidable computational challenge  \cite{MemMilPrz10}, which is 
usually overcome by looking at fit based on comparing network properties. Our approach gives a rigorous quantitative method for prioritising network properties; this is important as different properties might promote different potential models. 

The use of statistical mechanics to quantify the information content of network structure is well established; see e.g. 
\cite{AlbBar02,DorGolMen07,BiaCooVic08,AnnCooFerFraKle09}. Most work so far has focused on undirected networks. The network properties  most frequently studied are degree distributions, clustering coefficients, assortativities and path length statistics. There has also been research on occurrences of motifs and subgraphs, motivated by the idea that if a network favours specific local topological patterns then these might reflect common local processes.
A particular benefit of the approach followed here and in  \cite{AnnCooFerFraKle09} is the compact and explicit nature of the final formulae. 
Although their derivations are  involved in places, the final results are compact. They take easily measured topological observables as input, avoid the need for numerical simulations or approximations, and are easy and efficient to use as our (biological) datasets grow. We therefore  imagine that this line of research will continue to develop, by adding further macroscopic network observables, beyond degree statistics and degree correlation functions. Each addition will make the method more powerful and useful. 

The specific quantities calculated in this paper are: the Shannon entropy and complexity of directed graph ensembles with controlled degree distributions;  the Shannon entropy and complexity of directed graph ensembles with controlled degree distributions and controlled degree-degree correlation functions; and, the  
symmetrised Kullback-Leibler distance between pairs of such ensembles. For each of these we calculate the leading orders in the network size,  expressed in terms of the controlled degree distributions and degree-degree correlation functions of the ensembles concerned. 
We illustrate the use of our results in section  \ref{sec:results} with applications to experimental  data on gene regulation networks.

We adopt the following notation conventions. Each directed graph with $N$ nodes is defined by a matrix  $ \bc = \{c_{ij} \}$, with entries $c_{ij}\in\{0,1\}$ indicating whether ($c_{ij}=1$) or not ($c_{ij}=0$) there is a directed arc from node $j$ to node $i$.  For each node $i$ we define the so-called in- and out-degrees, viz. $k^{\rm out}_{i}(\bc)=\sum_j c_{ji}$ and  
$k^{\rm in}_{i}(\bc)=\sum_j c_{ij}$; in nondirected graphs such as in \cite{AnnCooFerFraKle09} one would have had $k_i^{\rm in}(\bc)=k_i^{\rm out}(\bc)$ for all $i$. We write the pair of degrees at a site $i$ as $\vec{k}_i(\bc)=(k_i^{\rm in}(\bc),k_i^{\rm out}(\bc))$.  Boldface letters will represent ordered sets with $N$ elements, such as $\bk^{\rm in} = (k^{in}_1, \ldots, k^{in}_N)$, or  $\bk^{\rm in}(\bc) = (k^{\rm in}_1(\bc), \ldots, k^{\rm in}_N(\bc))$.

\section{Directed graphs with controlled in- and out-degree distributions}
\label{sec:derivation_degree_only}

Here we calculate the Shannon entropy of an ensemble of directed random graphs constrained by a common joint distribution of in - and out-degrees. Via suitable adaptations of the methods developed for nondirected networks, we achieve a standard path-integral form to which we can apply the method of steepest descent. This leads to an elegant analytical expression for the entropy of the ensemble in the leading orders in $N$. The key term takes the form of a Kullback-Leibler distance between the imposed joint degree distribution and the Poissonnian one that would have been found upon generating directed arcs independently.

\subsection{Definition of the problem}
\label{sec:definitions}

We consider an ensemble of directed random graphs, where degree pairs  
	$\vec{k}_i=(k_i^{\rm in},k_i^{\rm out})$ are for each node $i$ drawn 
independently from a specified joint degree distribution
	$p(\vec{k})$:
	\begin{eqnarray}
		p(\bc ) &=& \sum_{\vec{k}_1\ldots\vec{k}_N} \Big[\prod_i p(\vec{k}_i)\Big]p(\bc|\vec{k}_1\ldots\vec{k}_N)
		\label{eq:p(c)}
\\
p(\bc|\vec{k}_1\!\ldots\vec{k}_N) &=& \frac{\prod_i \delta_{\vec{k}_i,\vec{k}_i(\bc)} }{Z(\vec{k}_1\!\ldots\vec{k}_N)},~~~~~~
 Z(\vec{k}_1\!\ldots\vec{k}_N) = \sum_{\bc} \prod_i \delta_{\vec{k}_i,\vec{k}_i(\bc)}  
		\label{eq:p(c|k)}
	\end{eqnarray}
For this ensemble we want to find the Shannon entropy per node $S=-N^{-1} \sum_{\bc} p(\bc)\log p(\bc)$, which informs us about the effective number 
${\cal N}=\exp(NS)$ of graphs in the ensemble and the complexity of directed graphs with the imposed degree statistics
	$p(\vec{k})$. 
Upon substituting  (\ref{eq:p(c|k)})  into the entropy formula, and after some simple manipulations and use of the law of large numbers, one finds that 
the entropy per node takes the form
	\begin{eqnarray}
	S &=& 
		\frac{1}{N}  
	 \! \! \sum_{\vec{k}_1\ldots\vec{k}_N}\!\! \Big[\prod_i p(\vec{k}_i)\Big]
                \log Z(\vec{k}_1\ldots\vec{k}_N)    
	- \!\sum_{\vec{k}} 
          		 p(\vec{k})
           		 \log p(\vec{k})
	 +   \epsilon_N
	\end{eqnarray}
where 
	$ \epsilon_N \rightarrow 0$ 
									as
											$ N \rightarrow \infty $.
To make the  first term in this expression more tractable, we transform $Z(\vec{k}_1\ldots\vec{k}_N)$ into an average
involving an alternative measure. If we denote the average degree by  $ \bar{k} = N^{-1}
\sum_i k^{\rm in}_i  = N^{-1}\sum_i k^{\rm out}_i $,  we may define the measure 
\begin{eqnarray}
w(\bc\vert  \bar{k}) &=& \prod_{ij} \Big[\frac{\bar{k}}{N} \delta_{c_{ij},1} \plus  \Big(1 \minus  \frac{\bar{k}}{N} \Big)\delta_{c_{ij},0} \Big]
\label{eq:poisson}
\nonumber
\\& =&
					\Big[1\minus \frac{\overline{k}}{N}
					\Big]^{N(N\minus 1)}
					\Big[\frac{\overline{k}/N}{1\! -\!\overline{k}/N}
					\Big]^{N\overline{k}(\bc)}
	\equiv~ W(\overline{k},\overline{k}(\bc))
	\end{eqnarray}
	Since this measure depends 
on the graph $\bc$ via $\bar{k}(\bc)$ only, we can  write the partition function
$Z(\vec{k}_1\ldots\vec{k}_N)$ in terms of an average over the measure (\ref{eq:poisson}), viz.   
	\begin{eqnarray}
	Z(\vec{k}_1\ldots\vec{k}_N)
&=&\frac{1}{W(\overline{k},\overline{k})}\sum_{\bc}w(\bc\vert \overline{k}) \prod_i \delta_{\vec{k}_i,\vec{k}_i(\bc)} 
	\end{eqnarray}
Introducing the notation $\bra f(\bc)\ket_\kappa=\sum_{\bc}w(\bc\vert \kappa)f(\bc)$ to represent 
averages over the measure (\ref{eq:poisson}) with average connectivity $\kappa$, the entropy per node can be written as
\begin{eqnarray}
S &=&  
	  \frac{1}{N}  
	 \! \! \sum_{\vec{k}_1\ldots\vec{k}_N}\!\! \Big[\prod_i p(\vec{k}_i)\Big]   
            \log \Big\bra \prod_i \delta_{\vec{k}_i,\vec{k}_i(\bc)}  \Big\ket_{\!\bar{k}} 
            - \sum_{\vec{k}}
                  p(\vec{k})  \log p(\vec{k}) 
             \nonumber
             \\
             &&
     -  \frac{1}{N}  
         \! \! \sum_{\vec{k}_1\ldots\vec{k}_N}\!\! \Big[\prod_i p(\vec{k}_i)\Big]   
                      \log
                      	\left[ \Big[1- \frac{\overline{k}}{N}\Big]^{N(N-1)}
								\Big[\frac{\overline{k}/N}{1-\overline{k}/N}\Big]^{N\overline{k}}   
						\right]             
 + \epsilon_N  
 \nonumber
 \\&=&  
   \frac{1}{N}  
	  \sum_{\vec{k}_1\ldots\vec{k}_N} \Big[\prod_i p(\vec{k}_i)\Big]   
            \log \Big\bra \prod_i \delta_{\vec{k}_i,\vec{k}_i(\bc)}  \Big\ket_{\!\bar{k}} 
            - \sum_{\vec{k}}
                  p(\vec{k})  \log p(\vec{k}) 
             \nonumber
             \\
             &&
    \hspace*{30mm} + \bra k\ket\big[ \log(N/\bra k\ket)     +1\big]                   
  + \varepsilon_N  
  \label{eq:S_before_Fourier}
	\end{eqnarray}
with $\lim_{N\to\infty}\varepsilon_N=0$, and with $\bra k\ket=\sum_{\vec{k}}k^{\rm in}p(\vec{k})=\sum_{\vec{k}}k^{\rm out}p(\vec{k})$. 	
All the complexity of the problem is thus contained in the first term of (\ref{eq:S_before_Fourier}):
	\begin{eqnarray}
	\phi &=&  
	 \frac{1}{N}  
	  \sum_{\vec{k}_1\ldots\vec{k}_N} \Big[\prod_i p(\vec{k}_i)\Big]   
            \log \Big\bra \prod_i \delta_{\vec{k}_i,\vec{k}_i(\bc)}  \Big\ket_{\!\bar{k}} 
             \label{eq:phi}
	\end{eqnarray}

\subsection{Entropy evaluation}

Using Fourier representations of the Kronecker deltas in (\ref{eq:phi}) and some straightforward manipulations brings us to
	\begin{eqnarray}
 && \phi=	\frac{1}{N}\!\!
					 \sum_{\vec{k}_1\ldots\vec{k}_N} \!\! \!
				 \Big[\prod_i p(\vec{k}_i)\Big]   
									\log \!	
										\int_{\minus \pi }^{\pi } \! 
											\prod_{i} 
						\Big[						\frac{\rmd\omega_i\rmd\psi_i}{4\pi^2}\rme^{\rmi[\omega_i k_i^{\rm in}+\psi_i k_i^{\rm out}]}	
						\Big]
												L(\bomega, \bpsi)
\\
&&
L(\bomega,\bpsi) =	\exp\Big[\bar{k}N\Big(\frac{1}{N}\!\sum_{i}\rme^{-\rmi\omega_i}\Big)\Big(\frac{1}{N}\!
	 \sum_j\rme^{-\rmi \psi_j}\Big)- \bar{k}N
						+  \mathcal{O}(N^{0})\Big]					
\end{eqnarray}
Introducing the quantities $R(\bomega)=N^{-1}\sum_i \rme^{-\rmi\omega_i}$ and $S(\bpsi)=N^{-1}\sum_i \rme^{-\rmi\psi_i}$, 
and inserting $\int\!\rmd R\rmd S~\delta[R-R(\bomega)]\delta[S-S(\bpsi)]$ with $\delta$-functions written in integral form, allows us to write 
\begin{eqnarray}
L(\bomega,\bpsi) &=& \int\!\frac{\rmd R\rmd\hat{R}\rmd S\rmd\hat{S}}{4\pi^2/N^2}\rme^{N\big[\rmi (\hat{R}R+\hat{S}S)
+\bar{k}(RS-1)\big]+  \mathcal{O}(N^{0})}\nonumber
\\
&&\times\prod_i \rme^{-\rmi[\hat{R}\rme^{-\rmi\omega_i}	+\hat{S}\rme^{-\rmi\psi_i}]}					
\end{eqnarray}
Substituting this back into $\phi$, using the law of large numbers, then gives
\begin{eqnarray}
\phi&=&	\frac{1}{N}\!\!
					 \sum_{\vec{k}_1\ldots\vec{k}_N} \!\!
				 \Big[\prod_i p(\vec{k}_i)\Big]   
									\log 	
									 \int\!\rmd R\rmd\hat{R}\rmd S\rmd\hat{S}~\rme^{N \Psi(R,\hat{R},S,\hat{S})+\order(\log N)}
									 \label{eq:saddle_point_1}
\end{eqnarray}
where 
\begin{eqnarray}
\hspace*{-10mm}
\Psi(R,\hat{R},S,\hat{S})&=& \rmi (\hat{R}R\!+\!\hat{S}S)
									 +\bar{k}(RS\!-\!1)									 
	+\sum_{k^{\rm in}}p(k^{\rm in}) \log\! 	\int_{\minus \pi }^{\pi } \! 
			\frac{\rmd\omega}{2\pi}\rme^{\rmi[\omega k^{\rm in}-\hat{R}\rme^{-\rmi\omega}]}	
			\nonumber
			\\
			\hspace*{-5mm}
			&&
			+\sum_{k^{\rm out}}p(k^{\rm out}) \log\!	\int_{\minus \pi }^{\pi } \! 
			\frac{\rmd\psi}{2\pi}\rme^{\rmi[\psi k^{\rm out}-\hat{S}\rme^{-\rmi\psi}]}
\end{eqnarray}
 The average in (\ref{eq:saddle_point_1}) over degree sequences is now obsolete since the argument depends in leading order in $N$ 
 on their distribution only, and 
(\ref{eq:saddle_point_1})  can be evaluated by steepest descent:
\begin{eqnarray}
\lim_{N\rightarrow \infty} 	\phi
											&=&		\extr_{R, \hat{R}, S, \hat{S}} \Psi(R, \hat{R}, S, \hat{S}]
\end{eqnarray}
We can simplify $\Psi$ by doing the remaining integrals, using 
\begin{eqnarray}
\int_{\minus \pi }^{\pi } \! 
			\frac{\rmd\omega}{2\pi}\rme^{\rmi[\omega k-A\rme^{-\rmi\omega}]}
			=\sum_{m\geq 0}\frac{(-\rmi A)^m}{m!}
			\int_{\minus \pi }^{\pi } \! 
			\frac{\rmd\omega}{2\pi}\rme^{\rmi\omega( k-m)}=\frac{(\!-\rmi A)^k}{k!}
\end{eqnarray}
Hence
\begin{eqnarray}
\Psi(R,\hat{R},S,\hat{S})&=& \rmi (\hat{R}R\!+\!\hat{S}S)
									 +\bar{k}(RS\!-\!1)									 
	+\sum_{k^{\rm in}}p(k^{\rm in}) \log [(-\rmi \hat{R})^{k^{\rm in}}\!\!/k^{\rm in}!]
	\nonumber
	\\&&
			+\sum_{k^{\rm out}}p(k^{\rm out}) \log [(-\rmi \hat{S})^{k^{\rm out}}\!\!/k^{\rm out}!]
\end{eqnarray}
Differentiation of $\Psi$ gives the following saddle-point equations:
\begin{eqnarray}
 -\rmi \hat{R} =\bar{k}S,~~~~~~~~	 -\rmi \hat{S} =\bar{k}R								 
\\	
			 \rmi R\hat{R}	+					 
	\bar{k}
		=0,~~~~~~ \rmi S\hat{S}	+									 
	\bar{k} =0
\end{eqnarray}
We conclude that $RS= 1$, and hence at the saddle-point we have
\begin{eqnarray}
\Psi(R,\hat{R},S,\hat{S})&=&
		\sum_{k^{\rm in}}p(k^{\rm in})\log \pi_{\bar{k}}(k^{\rm in})
							+\sum_{k^{\rm out}}p(k^{\rm out})\log\pi_{\bar{k}}(k^{\rm out})
							\label{eq:PSI_deg_only}
\end{eqnarray}	
with the Poissonnian degree distribution $\pi_{\bar{k}}(k)=\rme^{-\bar{k}}\bar{k}^k/k!$.

\subsection{Final analytical expression for the entropy of the ensemble}

The intermediate result (\ref{eq:PSI_deg_only}) can now be substituted back into the expression for the entropy of the constrained random graph ensemble defined in (\ref{eq:S_before_Fourier}), giving
\begin{eqnarray}
\label{eq:final_form_degree_only}
\hspace*{-10mm}
        S &=& \bar{k} 
        \big[ \log (N/\bar{k}) + 1\big]
     -                
          \! \! \sum_{k^{\rm in}\!, k^{\rm out}} \! \! 
          p(k^{\rm in}\!,k^{\rm out})
						\log 
						\Big( 
							\frac	{ p(k^{\rm in}\!,k^{\rm out})}
									{
										\pi_{\bar{k}}(k^{\rm in}) 
										\pi_{\bar{k}}(k^{\rm out}) 
									} 
						\Big)
		+  	\zeta_N
\end{eqnarray} 
where $\bar{k}$ is the average connectivity, $N$ is the number of nodes in the network, $p( k^{\rm in}\!,k^{\rm out})$ is its degree distribution that constrained the random graph ensemble, and $\lim_{N\to \infty}\zeta_N=0$. 

The compact form of (\ref{eq:final_form_degree_only}) enables us to interpret and understand this result for the entropy per node. 
For example, we can consider what the result would have been if the constraint on the ensemble had been less restrictive. If our ensemble was
a maximum entropy ensemble on the space of all directed graphs,  but now constrained by the average degree only (as opposed to the full joint in- and out-degree distribution), then the entropy per node would have been
$ S = \bar{k}[\log (N/\bar{k})+ 1]$. We see that this is identical to what we would obtain from (\ref{eq:final_form_degree_only}) if the constraining degree-distribution was 
 $p( k^{\rm in}\!, k^{\rm out}) = \pi(k^{\rm in}) \pi(k^{\rm out})$; a trivial calculation confirms that in the maximum entropy ensemble with constrained average degree one indeed has $p( k^{\rm in}\!, k^{\rm out}) = \pi(k^{\rm in}) \pi(k^{\rm out})$ for $N\to\infty$.
Similarly, if we had chosen a maximum entropy ensemble of directed graphs constrained by a prescribed degree sequence (as opposed to a joint degree distribution), then the entropy would have taken the form
\begin{eqnarray}
\hspace*{-10mm}
 S &=&
 \bar{k} 
 \big[\log (N/\bar{k})+1\big]
+  \! \! \sum_{k^{\rm in} \!, k^{\rm out}}
	p (k^{\rm in} \!, k^{\rm out})
	\log[ \pi(k^{\rm in}) \pi(k^{\rm out})]
		+  	\zeta_N	
		\label{eq:entropy_degreesonly}
\end{eqnarray}
This value is seen to be simply 
(\ref{eq:final_form_degree_only}) minus the Shannon entropy of the joint degree distribution $p(k^{\rm in}\!,k^{\rm out})$, 
reflecting the possible ways to relabel sites in the original ensemble; this freedom is removed once we specify the individual 
degrees rather than their distribution.

\section{Directed graphs with controlled degree distributions and degree-degree correlation functions}
\label{sec:ddc_derivation}

We extend our calculation to directed graph ensembles that are constrained further, by imposing 
a degree-degree correlation function in addition to a degree distribution. Degree-degree correlations in networks are known to carry 
valuable information. They can give rise to properties such as `assortativity' or `disassortativity' and often reflect the algorithm responsible for a network's generation. One such algorithm,  `preferential attachment',  is well illustrated by the World Wide Web, where pages are more likely to be `linked' to if they already have many pages linking to them. Preferential attachment models  such as \cite{AlbBar02} gained credibility by 
reproducing the typical fat tails often found in the degree distributions of real networks. 

\subsection{Definition of the problem}

We now wish to generate graphs with degree pairs $(k_i^{\rm in},k_i^{\rm out})$ again drawn independently from the distribution $p(\vec{k})=p(k^{\rm in}\!, k^{\rm out})$, but now the link probabilities are modified by some function $Q(\bki, \bkj|\bar{p})$ of the 
degrees of the nodes concerned, and their distribution, with $\bki =(k^{\rm in}_i, k^{\rm out}_i)$:
\begin{eqnarray}
p(\bc|p,Q) &=&  \sum_{\vec{k}_1\ldots \vec{k}_N}
	 \Big[\prod_i p(\vec{k}_i)\Big] p(\bc\vert \vec{k}_1\ldots\vec{k}_N,Q)
	\label{eq:fullfamily}
	\\
	 p(\bc\vert \vec{k}_1\ldots\vec{k}_N,Q)
&=&
 \frac{w(\bc|\vec{k}_1\ldots\vec{k}_N, Q) \prod_i \delta_{\vec{k}_i,\vec{k}_i(\bc)} }{Z(\vec{k}_1\ldots\vec{k}_N,Q)}
\\
Z(\vec{k}_1\ldots\vec{k}_N,Q)&=&  \sum_{\bc}w(\bc|\vec{k}_1\ldots\vec{k}_N, Q) \prod_i \delta_{\vec{k}_i,\vec{k}_i(\bc)} 
\nonumber
\end{eqnarray}
The difference with the graph ensemble in the previous section is the appearance of a 
new measure $w(\bc|\vec{k}_1\ldots\vec{k}_N, Q)$, defined as
\begin{eqnarray}
\hspace*{-10mm}
w(\bc\vert \vec{k}_1\ldots\vec{k}_N, Q)&=&
\prod_{i\neq j}\Big[\frac{\overline{k}}{N}Q(\bki,\bkj|\bar{p})\delta_{c_{ij},1}\plus 
\Big(1\minus \frac{\overline{k}}{N}Q(\bki,\bkj|\bar{p})\Big)\delta_{c_{ij},0}\Big]
\label{eq:deformation}
\end{eqnarray}
with $Q(\bki,\bkj|\bar{p})\geq 0$ for all $(\bki,\bkj)$, and with the distribution $\bar{p}(\vec{k})=N^{-1}\sum_i \delta_{\vec{k},\vec{k}_i}$
and the average degree $\overline{k}\!=\!N^{-1}\sum_i k_i^{\rm in}\!=\!N^{-1}\sum_i k_i^{\rm out}$ of the imposed degree sequence. 
The objective of the measure (\ref{eq:deformation})  is to deform the graph probabilities such as to impose a specific correlation profile 
between the degrees of connected nodes, by a suitable choice of the kernel $Q(.,.)$.
We take $Q(.,.)$ to be normalized such that $w(\bc|\ldots)$ is asymptotically consistent with the average degree $\bar{k}$. This means that we demand $N^{-2}\sum_{ij}Q(\vec{k}_i, \vec{k}_j|\bar{p})=1$. Equivalently, $\sum_{\vec{k},\vec{k}^\prime}\bar{p}(\vec{k})\bar{p}(\vec{k}^\prime)Q(\vec{k},\vec{k}^\prime|\bar{p})=1$, which explains why $Q(.,.)$ depends on the distribution $\bar{p}$. 
The entropy per node $S$ of our ensemble is
\begin{eqnarray}
&& S = -\sum_{\bc}p(\bc|p,Q)\Omega(\bc|p,Q)
\label{eq:SwithQ}
\\
&& \Omega(\bc|p,Q)= N^{-1}\log p(\bc|p,Q)
 \label{eq:Omega}
 \end{eqnarray}

 \subsection{Entropy evaluation}
 
 In \ref{app:OmegaCalculation} we calculate the quantity (\ref{eq:Omega}) in leading orders in $N$, resulting in formula (\ref{eq:Omega_result}). 
Substitution into expression (\ref{eq:SwithQ}) for the entropy, followed by doing the average over $p(\bc|p,Q)$ and some simple re-arranging of terms,  then gives us 
 \begin{eqnarray}
 S&=& 
 \bar{k}\big[\log(N/\overline{k})+1\big]
 -  \sum_{\vec{k}}p(\vec{k}) \log \Big[\frac{p(\vec{k})}{
 \pi_{\bar{k}}(k^{\rm in})\pi_{\bar{k}}(k^{\rm out})}\Big]
 \nonumber
\\
&&\hspace*{-5mm}
-
\bar{k}\sum_{\vec{k},\vec{k}^\prime}W(\vec{k},\vec{k}^\prime) \log \Big[\frac{R(\vec{k}|p,Q)Q(\vec{k},\vec{k}^\prime|p)S(\vec{k}^\prime|p,Q)}{W_1(\vec{k})W_2(\vec{k}^\prime)}\Big] 
 +  \tilde{\zeta}_N
 \label{eq:Snearly}
\end{eqnarray}
with $\lim_{N\to\infty}\tilde{\zeta}_N=0$, $\pi_{\bar{k}}(k)=\rme^{-\bar{k}}\bar{k}^k/k!$, and $\bar{k}=\sum_{\vec{k}}p(\vec{k})k^{\rm in}=\sum_{\vec{k}}p(\vec{k})k^{\rm out}$. 
The kernel $W(\vec{k},\vec{k}^\prime)$ and its two marginals $W_{1,2}(\vec{k})$ in this expression are as defined in 
(\ref{eq:defineW},\ref{eq:Wmarginal1},\ref{eq:Wmarginal2}), but now calculated for graphs from our ensemble 
(\ref{eq:fullfamily}). Similarly, the quantities $R(\vec{k}|p,Q)$ and $Q(\vec{k}|p,Q)$ are now solved from 
\begin{eqnarray}
\hspace*{-5mm}
 R(\vec{k})=\frac{ p(\vec{k})k^{\rm in}}
{\overline{k}\sum_{\vec{k}^\prime}Q(\vec{k},\vec{k}^\prime|p)S(\vec{k}^\prime)},
  ~~~~~~~~
S(\vec{k})= \frac{p(\vec{k})k^{\rm out}}
{ \overline{k}\sum_{\vec{k}^\prime}Q(\vec{k}^\prime,\vec{k}|p)R(\vec{k}^\prime)}
\label{eq:SQeqns}
\end{eqnarray}
in which the distribution $p(\vec{k})$, its associated average $\bar{k}$, as well as the kernel $Q(\vec{k},\vec{k}^\prime|p)$, correspond to 
ensemble (\ref{eq:fullfamily}). Thus the correct normalization of the kernel $Q(.,.)$ is $\sum_{\vec{k},\vec{k}^\prime}p(\vec{k})p(\vec{k})Q(\vec{k},\vec{k}^\prime|p)=1$. 
What remains is to express the distribution $W(\vec{k},\vec{k}^\prime|p,Q)$ for ensemble (\ref{eq:fullfamily})
in terms of $\{p,Q\}$. This is done in \ref{app:findW}, resulting in (\ref{eq:Wfound}): 
\begin{eqnarray}
\lim_{N\to\infty}
W(\vec{k},\vec{k}^\prime) &=& R(\vec{k}|p,Q) Q(\vec{k},\vec{k}^\prime|p)
	S(\vec{k}^\prime|p,Q)
\end{eqnarray}
in which $R(\vec{k}|p,Q)$ and $S(\vec{k}|p,Q)$ are once more the solutions of (\ref{eq:SQeqns}), but now with $\tilde{p}(\vec{k})$ replaced by $p(\vec{k})$. 
Combination with (\ref{eq:Snearly}) then gives us 
 \begin{eqnarray}
S &=&
\bar{k}[\log(N/\bar{k})+1]
-\sum_{\vec{k}}p(\vec{k})\log\Big[\frac{p(\vec{k})}{\pi_{\bar{k}}(k^{\rm in})\pi_{\bar{k}}(k^{\rm out})}\Big]
\nonumber
\\
&&
-\bar{k}
\sum_{\vec{k},\vec{k}^\prime}W(\vec{k},\vec{k}^\prime)
\log\Big[\frac{W(\vec{k},\vec{k}^\prime)}{W_1(\vec{k})W_2(\vec{k}^\prime)}\Big]
	+\tilde{\epsilon}_{N}
\label{eq:final_complex_entropy}
\end{eqnarray}
with $\lim_{N\to\infty}\tilde{\epsilon}_N=0$. 
Compared to the entropy per node  (\ref{eq:entropy_degreesonly}) of ensembles where only the in-out degree distributions are imposed, we see 
that imposing in addition our new constraint, the specific degree-degree correlations as embodied by $W(\vec{k},\vec{k}^\prime)$, leads to a reduction of the entropy by an amount proportional to the mutual information of in-out degrees of connected nodes. 
An analogous result was derived in \cite{AnnCooFerFraKle09} for nondirected graphs. 
 It can immediately be seen that if the in-out degrees of connected nodes are statistically independent, then the final nonvanishing term of 
 \ref{eq:final_complex_entropy}
will be zero. Hence the entropy of the ensemble will in that case be the same as though the only constraint was the degree distribution.  

\section{Quantifying structural distance between networks}
\label{sec:KL_Distance}

\subsection{Derivation of the distance formula}

In this section we define and calculate an information theoretic distance between two directed networks $A$ and $B$, with in-out degree distributions $p_A(\vec{k})$ and $p_B(\vec{k})$ and with degree-degree correlation functions $W_A(\vec{k},\vec{k}^\prime)$ and 
$W_B(\vec{k},\vec{k}^\prime)$. We generalize to the present context of directed graphs the choice made in \cite{AnnCooFerFraKle09}, viz. the Jeffreys divergence (i.e. symmetrized Kullback-Leibler distance) per node of the two associated ensembles from our family (\ref{eq:fullfamily}):
\begin{eqnarray}
D_{AB} 
&=&
\frac{1}{2N}
 \sum_{\bc} \Big\{
 p(\bc\vert p_A, \!  Q_A) \log\Big[ \frac{p(\bc\vert p_A, \! Q_A)}{p(\bc\vert p_B, Q_B)}\Big] 
 \nonumber
 \\
 &&\hspace*{10mm}+
p(\bc\vert p_B, Q_B) \log \Big[ \frac{p(\bc\vert p_B, Q_B)}{p(\bc|p_A, Q_A)} \Big]\Big\}
\label{eq:distance_defn}
\end{eqnarray}
$D_{AB}$ is non-negative and equals zero only when both networks $A$ and $B$ belong to the same tailored graph ensemble (i.e. have equivalent constraints). 
Upon writing the Shannon entropies per node of the ensembles $A$ and $B$ as $S_A$ and $S_B$, we have 
\begin{eqnarray}
D_{AB} 
&=&
\frac{1}{2}(S_{AB}+S_{BA}-S_{AA}-S_{BB})
\label{eq:Dformula}
\end{eqnarray}
where, using the abbreviation (\ref{eq:Omega}), 
 \begin{eqnarray}
 S_{AB}&=& 
- \frac{1}{N}\sum_{\bc}p(\bc\vert p_A, Q_A) \log p(\bc\vert p_B, Q_B)\nonumber
\\
&=& - \sum_{\bc}p(\bc\vert p_A, Q_A) \Omega(\bc|p_B,Q_B)
\label{eq:SAB}
\end{eqnarray}
with $\Omega(\bc|p,Q)$ as defined in  (\ref{eq:Omega}).
We may now use result  (\ref{eq:Omega_result}) 
of \ref{app:OmegaCalculation}, but in doing so it is vital to keep track carefully of the labels $(A,B)$ of the degree distributions and kernels. 
In particular, according to (\ref{eq:SAB}) we must make in  (\ref{eq:Omega_result}) the substitutions
$p(\vec{k}|\bc)\to p_A(\vec{k})$, $W(\vec{k},\vec{k}^\prime|\bc)\to W_A(\vec{k},\vec{k}^\prime)$, $p(\vec{k})\to p_B(\vec{k})$, and $Q(\vec{k},\vec{k}^\prime|\tilde{p})\to
 Q_B(\vec{k},\vec{k}^\prime|p_A)$. This leads us to
 \begin{eqnarray}
\hspace*{-20mm}
\lim_{N\to\infty}S_{AB}&=&	- \sum_{\vec{k}}p_A(\vec{k}) \log p_B(\vec{k})
-\bar{k}_A\Big[1\!+\!\log\big(\frac{\overline{k}_A}{N}\big)\Big]
 -\sum_{\vec{k}}p_A(\vec{k})
 \log (k^{\rm in}!k^{\rm out}!)
 \nonumber
 \\
 \hspace*{-20mm}
 &&\hspace*{-0mm}
 +\sum_{\vec{k}}p_A(\vec{k})k^{\rm in}
 \log \Big[ \frac{p_A(\vec{k})k^{\rm in}}{R(\vec{k}|p_A,Q_B)}\Big]
+\sum_{\vec{k}}p_A(\vec{k})k^{\rm out}\log 
  \Big[\frac{p_A(\vec{k})k^{\rm out}}{S(\vec{k}|p_A,Q_B)}\Big]
  \nonumber
\\
\hspace*{-20mm}
&&
\hspace*{10mm}
-
\bar{k}_A\sum_{\vec{k},\vec{k}^\prime}W_A(\vec{k},\vec{k}^\prime) \log Q_B(\vec{k},\vec{k}^\prime|p_A) 
\end{eqnarray}
in which $R(\vec{k}|p_A,Q_B)$ and $S(\vec{k}|p_A,Q_B)$ are to be solved from 
 \begin{eqnarray}
 \hspace*{-15mm}
 R(\vec{k})=\frac{ p_A(\vec{k})k^{\rm in}}
{\overline{k}_A\sum_{\vec{k}^\prime}\!Q_B(\vec{k},\vec{k}^\prime|p_A)S(\vec{k}^\prime)},
  ~~~~~~~
S(\vec{k})= \frac{p_A(\vec{k})k^{\rm out}}
{ \overline{k}_A\sum_{\vec{k}^\prime}\!Q_B(\vec{k}^\prime,\vec{k}|p_A)R(\vec{k}^\prime)}
\end{eqnarray}
Hence, upon assembling and combining  the various terms in (\ref{eq:Dformula}) 
and upon using relations such as (\ref{eq:Wmarginal1},\ref{eq:Wmarginal2}) and (\ref{eq:Wfound}) to simplify the result, we find 
\begin{eqnarray}
 \hspace*{-15mm}
D_{AB} 
&=&
\frac{1}{2}\sum_{\vec{k}}p_A(\vec{k}) \log \Big[\frac{p_A(\vec{k})}{p_B(\vec{k})}\Big]
+ \frac{1}{2}\sum_{\vec{k}}p_B(\vec{k}) \log \Big[\frac{p_B(\vec{k})}{p_A(\vec{k})}\Big]
\nonumber
\\
 \hspace*{-15mm}
 &&\hspace*{-0mm}
 +\frac{1}{2}\bar{k}_A \sum_{\vec{k},\vec{k}^\prime}W_{A}(\vec{k},\vec{k}^\prime) \log \Big[ \frac{W_A(\vec{k},\vec{k}^\prime)}{R(\vec{k}|p_A,Q_B)Q_B(\vec{k},\vec{k}^\prime|p_A) S(\vec{k}^\prime|p_A,Q_B)}
\Big]
\nonumber
\\
 \hspace*{-15mm}
 &&\hspace*{-0mm}
+\frac{1}{2}\bar{k}_B \sum_{\vec{k},\vec{k}^\prime}W_{B}(\vec{k},\vec{k}^\prime) \log \Big[ \frac{W_B(\vec{k},\vec{k}^\prime)}{R(\vec{k}|p_B,Q_A)Q_A(\vec{k},\vec{k}^\prime|p_B) S(\vec{k}^\prime|p_B,Q_A)}
\Big]
\end{eqnarray}
According to (\ref{eq:Wfound}), the product $W_{AB}(\vec{k},\vec{k}^\prime)=R(\vec{k}|p_A,Q_B) Q_B(\vec{k},\vec{k}^\prime|p_A) 
 S(\vec{k}^\prime|p_A,Q_B)$ equals the joint distribution of in- and out- degrees of connected nodes in an ensemble of the family (\ref{eq:fullfamily}) 
 that would have been obtained upon choosing the hybrid combination $\{p_A,Q_B\}$ of degree distribution and wiring kernel, where $Q_B$ is normalized according to $\sum_{\vec{k},\vec{k}^\prime}p_A(\vec{k}) p_A(\vec{k}^\prime)Q_B(\vec{k},\vec{k}^\prime|p_A)=1$.  Similarly, the product 
 $W_{BA}(\vec{k},\vec{k}^\prime)=R(\vec{k}|p_B,Q_A) Q_A(\vec{k},\vec{k}^\prime|p_B) 
 S(\vec{k}^\prime|p_B,Q_A)$ would have been obtained for the ensemble $\{p_B,Q_A\}$. Thus we may write
 \begin{eqnarray}
\lim_{N\to\infty}D_{AB} 
&=&
  \frac{1}{2}\sum_{\vec{k}}p_A(\vec{k}) \log \Big[\frac{p_A(\vec{k})}{p_B(\vec{k})}\Big] +\frac{1}{2}
   \sum_{\vec{k}}p_B(\vec{k}) \log \Big[\frac{p_B(\vec{k})}{p_A(\vec{k})}\Big]
 \nonumber
 \\
 &&
 +
\frac{1}{2}\bar{k}_A\sum_{\vec{k},\vec{k}^\prime}W_A(\vec{k},\vec{k}^\prime) \log\Big[\frac{W_A(\vec{k},\vec{k}^\prime) }
{ W_{AB}(\vec{k},\vec{k}^\prime) }
 \Big]
   \nonumber
    \\
 &&
 +
\frac{1}{2}\bar{k}_B\sum_{\vec{k},\vec{k}^\prime}W_B(\vec{k},\vec{k}^\prime) \log \Big[ \frac{W_B(\vec{k},\vec{k}^\prime)}{W_{BA}(\vec{k},\vec{k}^\prime)}
 \Big]
 \label{eq:first_distance}
\end{eqnarray}
This appealing formula shows that $D_{AB}\geq 0$ for all choices of $(A,B)$, with equality if and only if $W_A=W_B$; 
in the later case one automatically will have $W_{AB}=W_{BA}=W_A=W_B$. 
In the case where degree-degree correlations are absent from both networks one will find 
$W_{AB}(\vec{k},\vec{k}^\prime) =W_{A}(\vec{k},\vec{k}^\prime) =W_{1A}(\vec{k})W_{2A}(\vec{k}^\prime)$, and formula 
(\ref{eq:first_distance}) reduces to the Jeffreys divergence between the degree distributions $p_A$ and $p_B$.

\subsection{Practical form of the distance formula}

In contrast to $W_A$ and $W_B$, which correspond to the two given networks $\bc_A$ and $\bc_B$, 
we cannot measure $W_{AB}$ and $W_{BA}$; the later would correspond to hypothetical hybrid networks.
Hence in order to use (\ref{eq:first_distance})  in practice it will be convenient to write it in an alternative form:
\begin{eqnarray}
 \hspace*{-15mm}
\lim_{N\to\infty}D_{AB} 
&=&
  \frac{1}{2}\sum_{\vec{k}}p_A(\vec{k}) \log \Big[\frac{p_A(\vec{k})}{p_B(\vec{k})}\Big] +\frac{1}{2}
   \sum_{\vec{k}}p_B(\vec{k}) \log \Big[\frac{p_B(\vec{k})}{p_A(\vec{k})}\Big]
 \nonumber
 \\
  \hspace*{-15mm}
 &&
 \hspace*{-15mm}
 +
\frac{1}{2}\bar{k}_A\sum_{\vec{k},\vec{k}^\prime}W_A(\vec{k},\vec{k}^\prime) \log\Big[\frac{W_A(\vec{k},\vec{k}^\prime) }
{ W_{B}(\vec{k},\vec{k}^\prime) }
 \Big]
 +
\frac{1}{2}\bar{k}_B\sum_{\vec{k},\vec{k}^\prime}W_B(\vec{k},\vec{k}^\prime) \log \Big[ \frac{W_B(\vec{k},\vec{k}^\prime)}{W_{A}(\vec{k},\vec{k}^\prime)}
 \Big]
 \nonumber
 \\
  \hspace*{-15mm}
 &&
 \hspace*{-10mm}
  +
\frac{1}{2}\bar{k}_A\sum_{\vec{k},\vec{k}^\prime}W_A(\vec{k},\vec{k}^\prime) \log\Big[\frac{W_B(\vec{k},\vec{k}^\prime)
 }
{ R(\vec{k}|p_A,Q_B) Q_B(\vec{k},\vec{k}^\prime|p_A)S(\vec{k}^\prime|p_A,Q_B)}
 \Big]
   \nonumber
    \\
     \hspace*{-15mm}
 &&
 \hspace*{-10mm}
  +
\frac{1}{2}\bar{k}_B\sum_{\vec{k},\vec{k}^\prime}W_B(\vec{k},\vec{k}^\prime) \log \Big[ \frac{W_A(\vec{k},\vec{k}^\prime)}{
R(\vec{k}|p_B,Q_A) Q_A(\vec{k},\vec{k}^\prime|p_B)S(\vec{k}^\prime|p_B,Q_A)}
 \Big]
 \label{eq:intermediateD}
\end{eqnarray}
If we choose $Q_A$ and $Q_B$ to be the canonical kernels for the two ensembles $A$ and $B$, i.e. 
$Q_A(\vec{k},\vec{k}^\prime|\bar{p})=W_A(\vec{k},\vec{k}^\prime)/\bar{p}(\vec{k}) \bar{p}(\vec{k}^\prime)$ and $Q_B(\vec{k},\vec{k}^\prime|\bar{p})=W_B(\vec{k},\vec{k}^\prime)/\bar{p}(\vec{k}) \bar{p}(\vec{k}^\prime)$, expression (\ref{eq:intermediateD}) 
simplifies to 
\begin{eqnarray}
 \hspace*{-15mm}
\lim_{N\to\infty}D_{AB} 
&=&
  \frac{1}{2}\sum_{\vec{k}}p_A(\vec{k}) \log \Big[\frac{p_A(\vec{k})}{p_B(\vec{k})}\Big] +\frac{1}{2}
   \sum_{\vec{k}}p_B(\vec{k}) \log \Big[\frac{p_B(\vec{k})}{p_A(\vec{k})}\Big]
 \nonumber
 \\
  \hspace*{-15mm}
 &&
 \hspace*{-15mm}
 +
\frac{1}{2}\bar{k}_A
\sum_{\vec{k},\vec{k}^\prime}W_A(\vec{k},\vec{k}^\prime) \log\Big[\frac{W_A(\vec{k},\vec{k}^\prime) }
{ W_{B}(\vec{k},\vec{k}^\prime) }
 \Big]
 +
\frac{1}{2}\bar{k}_B\sum_{\vec{k},\vec{k}^\prime}W_B(\vec{k},\vec{k}^\prime) \log \Big[ \frac{W_B(\vec{k},\vec{k}^\prime)}{W_{A}(\vec{k},\vec{k}^\prime)}
 \Big]
 \nonumber
 \\
  \hspace*{-15mm}
 &&
 \hspace*{-16mm}
  +\frac{1}{2}\bar{k}_A\Big\{\sum_{\vec{k}}W_{1A}(\vec{k}) \log\Big[\frac{p_A(\vec{k}) }{ R(\vec{k}|p_A,Q_B) }
 \Big]
 +
\sum_{\vec{k}^\prime}W_{2A}(\vec{k}^\prime) \log\Big[\frac{p_A(\vec{k}^\prime)}{ S(\vec{k}^\prime|p_A,Q_B)}
 \Big]\Big\}
   \nonumber
    \\
     \hspace*{-15mm}
 &&
 \hspace*{-16mm}
  +
\frac{1}{2}\bar{k}_B\Big\{
\sum_{\vec{k}}W_{1B}(\vec{k}) \log \Big[ \frac{p_B(\vec{k})}{
R(\vec{k}|p_B,Q_A) }
 \Big]
 + 
 \sum_{\vec{k}^\prime}W_{2B}(\vec{k}^\prime) \log \Big[ \frac{p_B(\vec{k}^\prime)}{
S(\vec{k}^\prime|p_B,Q_A)}
 \Big]
 \Big\}
 \nonumber
 \\[-1mm]
 \hspace*{-15mm}&&
 \label{eq:intermediateD2}
\end{eqnarray}
with $R(\vec{k}|p_A,Q_B) $ and $S(\vec{k}|p_A,Q_B) $ to be solved from 
 \begin{eqnarray}
 R(\vec{k})/p_A(\vec{k})=\frac{W_{1A}(\vec{k})}
{\sum_{\vec{k}^\prime}\!W_B(\vec{k},\vec{k}^\prime)[S(\vec{k}^\prime)/p_A(\vec{k}^\prime)]},
\\
S(\vec{k})/p_A(\vec{k}^\prime)= \frac{W_{2A}(\vec{k})}
{ \sum_{\vec{k}^\prime}\!W_B(\vec{k}^\prime,\vec{k})[R(\vec{k}^\prime)/p_A(\vec{k}^\prime)]}
\end{eqnarray}
Next we rewrite the arguments of the logarithms 
in the second line of  (\ref{eq:intermediateD}) in terms of the two degree correlation ratios $\Pi_A(\vec{k},\vec{k}^\prime)=
W_A(\vec{k},\vec{k}^\prime)/W_{1A}(\vec{k})W_{2A}(\vec{k}^\prime)$ and 
$\Pi_B(\vec{k},\vec{k}^\prime)=
W_B(\vec{k},\vec{k}^\prime)/W_{1B}(\vec{k})W_{2B}(\vec{k}^\prime)$. We also transform the order parameters $R(\vec{k}|p_A,Q_B)$ and $S(\vec{k}|p_A,Q_B)$ to new functions $\rho_{AB}(\vec{k})$ and $\sigma_{AB}(\vec{k})$ via
\begin{eqnarray}
\hspace*{-10mm}
&&
\rho_{AB}(\vec{k})=\frac{p_A(\vec{k})W_{1A}(\vec{k})}{ R(\vec{k}|p_A,Q_B)W_{1B}(\vec{k})}
,~~~~~~\sigma_{AB}(\vec{k})=\frac{p_A(\vec{k})W_{2A}(\vec{k})}{S(\vec{k}|p_A,Q_B)W_{2B}(\vec{k})}
\end{eqnarray}
 Our distance then becomes
\begin{eqnarray}
 \hspace*{-15mm}
\lim_{N\to\infty}D_{AB} 
&=&
  \frac{1}{2}\sum_{\vec{k}}p_A(\vec{k}) \log \Big[\frac{p_A(\vec{k})}{p_B(\vec{k})}\Big] +\frac{1}{2}
   \sum_{\vec{k}}p_B(\vec{k}) \log \Big[\frac{p_B(\vec{k})}{p_A(\vec{k})}\Big]
 \nonumber
 \\
  \hspace*{-15mm}
 &&
 \hspace*{-13mm}
 +
\frac{1}{2}\bar{k}_A\sum_{\vec{k},\vec{k}^\prime}W_A(\vec{k},\vec{k}^\prime) \log\Big[\frac{\Pi_A(\vec{k},\vec{k}^\prime) }
{ \Pi_{B}(\vec{k},\vec{k}^\prime) }
 \Big]
 +
\frac{1}{2}\bar{k}_B\sum_{\vec{k},\vec{k}^\prime}W_B(\vec{k},\vec{k}^\prime) 
\log \Big[ \frac{\Pi_B(\vec{k},\vec{k}^\prime)}{\Pi_{A}(\vec{k},\vec{k}^\prime)}
 \Big]
 \nonumber
 \\
  \hspace*{-15mm}
 &&
   \hspace*{-10mm}
  +
\frac{1}{2}\bar{k}_A\sum_{\vec{k}}W_{1A}(\vec{k}) \log\rho_{AB}(\vec{k})
  +
\frac{1}{2}\bar{k}_A\sum_{\vec{k}}W_{2A}(\vec{k}) \log\sigma_{AB}(\vec{k})
  \nonumber
 \\
  \hspace*{-15mm}
 &&
   \hspace*{-10mm}
  +
\frac{1}{2}\bar{k}_B\sum_{\vec{k}}W_{1B}(\vec{k}) \log\rho_{BA}(\vec{k})  +
\frac{1}{2}\bar{k}_B\sum_{\vec{k}}W_{2B}(\vec{k}) \log\sigma_{BA}(\vec{k})
\label{eq:distance_transformed}
\end{eqnarray}
in which $\rho_{AB}(\vec{k})$ and $\sigma_{AB}(\vec{k})$ are to be solved from 
 \begin{eqnarray}
\rho_{AB}(\vec{k})&=&\sum_{\vec{k}^\prime}\Pi_B(\vec{k},\vec{k}^\prime)W_{2A}(\vec{k}^\prime) \sigma^{-1}_{AB}(\vec{k}^\prime)
\label{eq:rho}
\\
\sigma_{AB}(\vec{k})&=& \sum_{\vec{k}^\prime}\Pi_B(\vec{k}^\prime,\vec{k})W_{1A}(\vec{k}^\prime)\rho^{-1}_{AB}(\vec{k}^\prime)
\label{eq:sigma}
\end{eqnarray}
Whenever $p_A=p_B$ or $\Pi_A=\Pi_B$ (or both), the solution of (\ref{eq:rho},\ref{eq:sigma}) will be $\rho_{AB}(\vec{k})=\sigma_{AB}(\vec{k})=1$ for all $\vec{k}$. 
Hence the last two lines of (\ref{eq:distance_transformed}) represent corrections to the distance formula, that reflect interference 
between the constraints imposed by prescribed degree statistics and those imposed by presecribed degree correlations\footnote{A similar interference term was erroneously omitted from 
\cite{AnnCooFerFraKle09}, which can be confirmed by retracing the above arguments and the calculations in 
\ref{app:OmegaCalculation} for nondirected graphs. We will summarize and compare our results for directed and nondirected graphs below.}.

We note, finally, that 
although definition (\ref{eq:distance_defn}) requires that the networks $A$ and $B$ have the same number of nodes, the final form (\ref{eq:distance_transformed}) of our formula does not depend on the (relative) network sizes. Hence we will apply  the result  (\ref{eq:distance_defn}) also  to networks of different sizes, provided both are sufficiently large, which makes (\ref{eq:distance_defn})  more widely applicable to real networks (which will in general be large, but of different sizes).

\section{Tests, comparisons, and applications}
\label{sec:results}

\subsection{Simple special cases}

If the in-degrees are statistically independent of the out-degrees, i.e. 
$p(\vec{k})= p(k^{in}) p(k^{out})$, the entropy per node (\ref{eq:final_form_degree_only}) 
of the ensemble (\ref{eq:p(c)}) 
 with prescribed degree statistics but no degree correlations simplifies to 
\begin{eqnarray}
        S &=& \bar{k} 
        \big[ \log (\frac{N}{\bar{k}}) \!+\! 1\big]
     -                 \sum_{k^{\rm in}}  p(k^{\rm in})
						\log 
						\Big[\frac	{ p(k^{\rm in})}{\pi_{\bar{k}}(k^{\rm in}) }\Big]
						\nonumber
\\
&&\hspace*{20mm} - \sum_{k^{\rm out}} 
          p(k^{\rm out})
						\log 
						\Big[ 
							\frac	{ p(k^{\rm out})}
									{\pi_{\bar{k}}(k^{\rm out}) 
									} 
						\Big]
		+  	\zeta_N
\end{eqnarray} 
with $\lim_{N\to\infty}\zeta_N=0$. This, according to \cite{AnnCooFerFraKle09},  is the sum of the individual entropies of the `out-graph' ensemble and the `in-graph' ensemble, calculated as though they were considered as two separate undirected networks. In ensembles with degree correlations, i.e. (\ref{eq:fullfamily}), 
with entropy per node (\ref{eq:final_complex_entropy}), the additional term that represents the entropy reduction imposed by the degree correlations
does not simplify as a result of assuming $p(\vec{k})= p(k^{in}) p(k^{out})$; the degree correlations can generate statistical relations between  in- and out-degrees that are not visible in $p(\vec{k})$.

A regular directed graph is one where each node has the same in- and the same out-degree. Since for a well-defined directed graph, we also have  $\sum_{\vec{k}}p(\vec{k})k^{\rm in}=\sum_{\vec{k}}p(\vec{k})k^{\rm out}=\overline{k}$, any regular directed graph must have $p(\vec{k})=\delta_{\vec{k},(\overline{k},\overline{k})}$. This, in turn, implies also that $W(\vec{k},\vec{k}^\prime)=\delta_{\vec{k},(\overline{k},\overline{k})}\delta_{\vec{k}^\prime,(\overline{k},\overline{k})}$. So it is impossible to have degree correlations, and both equation (\ref{eq:final_form_degree_only}) and (\ref{eq:final_complex_entropy}) 
 reduce to
 \begin{eqnarray}
        S &=& \bar{k} 
        \big[ \log (N\bar{k}) -1\big]
     - 2 \log (\bar{k}!) 
		+  	\zeta_N
\end{eqnarray}

\subsection{Comparison of formulae for undirected versus directed networks}

It is instructive to give an overview of the similarities and differences between directed and nondirected graphs. Instead of entropies per node, 
we will also compare entropic results in terms of complexities. The degree complexity per node ${\cal C}_{\rm deg}$ of a graph $\bc$ is the difference between the entropy per node of the associated ensemble (\ref{eq:p(c)}) and the value $S_0[\bar{k}]$ that is found for the entropy per node if only the average connectivity $\bar{k}$ is prescribed (i.e. for an ensemble with Poisson distributed degrees). The wiring complexity ${\cal C}_{\rm wir}$ is the further entropy reduction that results if we go from the ensemble (\ref{eq:p(c)}) to the ensemble 
(\ref{eq:fullfamily}) where also the degree-degree correlations are imposed. 
Our results can then be summarized as in table \ref{tab:compare}.
\begin{table}[t]
\begin{eqnarray*}
\hspace*{-25mm}
& \underline{\large\sl directed~graphs} & \underline{\large\sl nondirected~graphs}
\\[3mm]
\hspace*{-25mm}
S_0[\bar{k}]: & \bar{k}\big[\log(N/\bar{k})+1\big]  & \frac{1}{2}\bar{k}\big[\log(N/\bar{k})+1\big]
\\[1mm]
\hspace*{-25mm}
{\cal C}_{\rm deg}[p]:~~~~ & \sum_{\vec{k}}p(\vec{k})\log\Big[\frac{p(\vec{k})}{\pi_{\bar{k}}(k^{\rm in})\pi_{\bar{k}}(k^{\rm out})}\Big]~~~~~~~~
& \sum_{k}p(k)\log\Big[\frac{p(k)}{\pi_{\bar{k}}(k)}\Big]
\\[1mm]
\hspace*{-25mm}
{\cal C}_{\rm wir}[p,W]:~~~~~ & \bar{k}
\sum_{\vec{k},\vec{k}^\prime}W(\vec{k},\vec{k}^\prime)
\log\Big[\frac{W(\vec{k},\vec{k}^\prime)}{W_1(\vec{k})W_2(\vec{k}^\prime)}\Big]
~~~~~~
&
\frac{1}{2}\bar{k}
\sum_{k,k^\prime}W(k,k^\prime)
\log\Big[\frac{W(k,k^\prime)}{W(k)W(k^\prime)}\Big]
\end{eqnarray*}
\caption{Comparison of entropies and complexities of directed versus nondirected graphs.
The entropy per node is given by $S[p,W]=S_0[\bar{k}]-{\cal C}_{\rm deg}[p]-{\cal C}_{\rm wir}[p,W]$, modulo finite size corrections. 
For ensembles in which only the average connectivity $\bar{k}$ is prescribed one would find the value $S_0[\bar{k}]$. The quantities ${\cal C}_{\rm deg}[p]$ and ${\cal C}_{\rm wir}[p,W]$ measure the entropy reductions caused by subsequently imposing a degree distribution $p$, and the joint distribution $W$ of connected nodes, and can therefore be identified with the degree complexity and the wiring complexity of the typical graphs in our ensembles. In directed graphs $\vec{k}=(k^{\rm in},k^{\rm out})$,  where $k_i^{\rm in}(\bc)=\sum_j c_{ij}$ and $k_i^{\rm out}(\bc)=\sum_j c_{ji}$, and $W(\vec{k},\vec{k}^\prime)=(N\bar{k})^{-1}\sum_{ij}c_{ij}\delta_{\vec{k},\vec{k}_i}
\delta_{\vec{k}^\prime,\vec{k}_j}$. In nondirected graphs one has only $k_i(\bc)=\sum_j c_{ij}$, and $W(k,k^\prime)
=(N\bar{k})^{-1}\sum_{ij}c_{ij}\delta_{k,k_i}
\delta_{k^\prime,k_j}$. }
\label{tab:compare}
\end{table}

\begin{table}[h]
\begin{eqnarray*}
\hspace*{-25mm}
&~~~ \underline{\large\sl directed~graphs} &~~~ \underline{\large\sl nondirected~graphs}
\\[3mm]
\hspace*{-25mm}
D_{AB}^{\rm deg}: 
   &~~~ \frac{1}{2}\sum_{\vec{k}}p_A(\vec{k}) \log \Big[\frac{p_A(\vec{k})}{p_B(\vec{k})}\Big] 
  &~~~ \frac{1}{2}\sum_{k}p_A(k) \log \Big[\frac{p_A(k)}{p_B(k)}\Big]
 \\[-1mm]
\hspace*{-25mm}  &~~~ +\frac{1}{2}
   \sum_{\vec{k}}p_B(\vec{k}) \log \Big[\frac{p_B(\vec{k})}{p_A(\vec{k})}\Big]
&~~~ +\frac{1}{2}
   \sum_{k}p_B(k) \log \Big[\frac{p_B(k)}{p_A(k)}\Big]
\\[2mm]
\hspace*{-25mm}
D_{AB}^{\rm wir}:~~ &  
\frac{1}{2}\bar{k}_A\sum_{\vec{k},\vec{k}^\prime}W_A(\vec{k},\vec{k}^\prime) \log\Big[\frac{\Pi_A(\vec{k},\vec{k}^\prime) }
{ \Pi_{B}(\vec{k},\vec{k}^\prime) }
 \Big]
&
\frac{1}{4}\bar{k}_A\sum_{k,k^\prime}W_A(k,k^\prime) \log\Big[\frac{\Pi_A(k,k^\prime) }
{ \Pi_{B}(k,k^\prime) }
 \Big] 
 \\[-1mm]
 \hspace*{-25mm}
 & 
 +
\frac{1}{2}\bar{k}_B\sum_{\vec{k},\vec{k}^\prime}W_B(\vec{k},\vec{k}^\prime) 
\log \Big[ \frac{\Pi_B(\vec{k},\vec{k}^\prime)}{\Pi_{A}(\vec{k},\vec{k}^\prime)}
 \Big]
 & 
 +
\frac{1}{4}\bar{k}_B\sum_{k,k^\prime}W_B(k,k^\prime) 
\log \Big[ \frac{\Pi_B(k,k^\prime)}{\Pi_{A}(k,k^\prime)}
 \Big]
\\[2mm]
\hspace*{-25mm}
D_{AB}^{\rm int}:~~~ &  
 \frac{1}{2}\bar{k}_A\sum_{\vec{k},\vec{k}^\prime}W_{A}(\vec{k},\vec{k}^\prime) \log[\rho_{AB}(\vec{k})
 \sigma_{AB}(\vec{k}^\prime)]
 & 
~~~ \frac{1}{2}\bar{k}_A\sum_{k}W_{A}(k) \log \rho_{AB}(k)
 \\[-1mm]
 \hspace*{-25mm}
 & +
\frac{1}{2}\bar{k}_B\sum_{\vec{k},\vec{k}^\prime}W_{B}(\vec{k},\vec{k}^\prime) \log[\rho_{BA}(\vec{k}) \sigma_{BA}(\vec{k}^\prime)]~~~
&
~~~+ \frac{1}{2}\bar{k}_B\sum_{k}W_{B}(k) \log \rho_{BA}(k)
\end{eqnarray*}
\caption{Comparison of the contributions to the distance
$\lim_{N\to\infty}D_{AB}=D_{AB}^{\rm deg}+D_{AB}^{\rm wir}+D_{AB}^{\rm int}$, between graphs $\bc_A$ and $\bc_B$.  Notation 
conventions are mostly as in the caption of table \ref{tab:compare}. The degree correlation ratios $\Pi$ are defined as
$\Pi(\vec{k},\vec{k}^\prime)=W(\vec{k},\vec{k}^\prime)/W_1(\vec{k})W_2(\vec{k}^\prime)$ (for directed graphs) 
and $\Pi(k,k^\prime)=W(\vec{k},\vec{k}^\prime)/W(k)W(k^\prime)$ (for nondirected graphs). The functions $\rho_{AB}(\vec{k})$ and 
$\sigma_{AB}(\vec{k})$ (for directed graphs) are the solutions of equations (\ref{eq:rho},\ref{eq:sigma}). The functions $\rho_{AB}(k)$ (for nondirected graphs) are to be solved from 
equation (\ref{eq:rho_nondirected}). }
\label{tab:compare_dist}.
\end{table}
Similarly we can compare the formulae for the information-theoretic distance $D_{AB}$ between two networks $\bc_A$ and $\bc_B$, for directed versus nondirected ones. This gives in both cases $\lim_{N\to\infty}D_{AB}=D_{AB}^{\rm deg}+D_{AB}^{\rm wir}+D_{AB}^{\rm int}$, where $D_{AB}^{\rm deg}$ 
is the direct contribution from degree distribution dissimilarity, $D_{AB}^{\rm wir}$ is the direct contribution from degree-correlation dissimilarity, and 
$D_{AB}^{\rm int}$ accounts for the interference between degree statistics and the possible degree correlations that could be achieved. 
Our distance results can then be summarized in table \ref{tab:compare_dist}.

The functions  $\rho_{AB}(\vec{k})$ and $\sigma_{AB}(\vec{k})$ are solved from  
(\ref{eq:rho},\ref{eq:sigma}). Repeating the calculation for nondirected graphs shows that there only one function
$\rho_{AB}(k)$ is required (or equivalently, $\rho_{AB}=\sigma_{AB}$), which is the solution of 
 \begin{eqnarray}
\rho_{AB}(k)&=&\sum_{k^\prime}\Pi_B(k,k^\prime)W_{A}(k^\prime) \rho^{-1}_{AB}(k^\prime)
\label{eq:rho_nondirected}
\end{eqnarray}

\subsection{Application to gene regulation networks}

A gene regulation network can be viewed as a directed graph, where the nodes represent genes and the arcs indicate whether ($c_{ij}=1$) 
or not ($c_{ij}=0$) the protein synthesized from gene $j$ acts as a regulator of gene $i$. 
In the present binary set-up, where $c_{ij}\in\{0,1\}$, one disregards information on the nature of regulation, i.e. whether it involves repression or activation. 

In tables \ref{tab:hughes} and \ref{tab:harbison} we show the results of calculating the various contributions to the entropy of the ensemble associated with the networks of 
\cite{citeulike:124} and \cite{citeulike:466068} respectively.  Imposing only the correct average degree gives the entropy $S_0[\bar{k}]$. Imposing in addition the correct degree distribution 
(i.e. representing the network by ensemble (\ref{eq:p(c)})) gives the entropy $S_0[\bar{k}]-{\cal C}_{\rm deg}[p]$. Imposing additionally the correct degree-degree correlations (i.e. representing the network by ensemble (\ref{eq:fullfamily})) reduces the entropy still further to  $S_0[\bar{k}]
-{\cal C}_{\rm deg}[p]-{\cal C}_{\rm wir}[p,W]$. 

In both tables  we also show the entropies per arc, defined as $S^\prime=S/\bar{k}$.
The latter are normalised for the average degree. This fits in with the `arc centric' view that the calculations in this paper and its predecessor \cite{AnnCooFerFraKle09}  seem to have steered us in, where  the final answers are consistently found to be most 
elegantly formulated in terms of the joint distribution $W$ of degrees at either end of an arc. 

In \cite{citeulike:124} Hughes \textit{et al.} used a two-color cDNA micro-array hybridization assay to generate expression profiles in yeast for  276 deletion mutants. We followed an approach published by Rung \textit{et al.} \cite{citeulike:2904850} to construct a network from this data. Two genes g1, g2 are connected by an arc from g1 to g2 if the ratio of the expression level  in the mutant where gene g1 is deleted versus the background standard deviation in the wild-type strain is larger than a threshold. In this way, we arrived at a directed network with $N=5654$ nodes (genes), with an average degree $\bar{k}\approx  5.6$. The degree distribution of this network is characterised by high frequency of occurrence of low degree nodes; the set of nodes with out-degree zero and in-degree less than 4 covers more than 50\% of the set. However, the network also contains some nodes with very high out-degree. 

\begin{table}[h]
\centering\vspace*{1mm}

\begin{tabular}{c}
{\em Gene regulation network of Hughes \textit{et al.}  (2000)}\\[2mm]
\begin{tabular}{ | l | c |c| }
\hline			
 Imposed topological property  & Entropy per node & Entropy per arc \\
\hline

\hline
average degree $\bar{k}\room$ & 44.5 & 7.9 \\
\hline
degree distribution $p(\vec{k})\room$ & 19.5 & 3.5 \\
\hline
degree-degree correlations $\Pi(\vec{k},\vec{k}^\prime)\room$ & 17.9 & 3.2 \\
\hline
\end{tabular}
\end{tabular}
\vsp

\caption{The tailoring of random graph ensembles by imposing as constraints the values of increasingly prescriptive macroscopic 
topological features measured in the gene regulation network of \cite{citeulike:124}. This tailoring reduces the entropy per node $S$ in the ensemble
in stages, and thereby the effective number of graphs 
${\cal N}=\exp[NS]$ compatible with the network of \cite{citeulike:124}. 
We observe that, in this example, refining the tailoring of the graph ensemble 
from imposing only the correct average degree 
to imposing the correct degree distribution is  more significant than the further refinement of imposing the correct degree-degree correlations. 
Hence the degree complexity of this network is significantly larger than the wiring complexity. }
\label{tab:hughes}
\end{table}

The authors of \cite{citeulike:466068}, Harbison \textit{et al.} reported on a study of DNA binding transcriptional regulators in yeast. For each of the 203 transcription factors tested they report the genes where the transcription factor bound to the putative promoter region. Similar to a previous study \cite{citeulike:307477} we constructed a network by connecting gene g1, which encodes a transcription factor,  to gene g2 if the measurements were statistically significant ($P\leq 0.001$).  Their data were represented as a directed network of $N=3865$ nodes, with an average degree of $\bar{k}\approx 2.81$. Compared with the  data of \cite{citeulike:124}, the network of \cite{citeulike:466068} is more sparse. It does, however, show a similar degree distribution pattern - in fact over 50\% of the nodes have zero out-degree and an in-degree of less than 2. 

\begin{table}[h]
\centering\vspace*{1mm}

\begin{tabular}{c}
{\em Gene regulation network of Harbison \textit{et al.}  (2004)}\\[2mm]
\begin{tabular}{ | l | c |c| }
\hline			
 Imposed topological property  & Entropy per node & Entropy per arc \\
\hline

\hline
average degree $\bar{k}\room$ & 23.2 & 8.2 \\
\hline
degree distribution $p(\vec{k})\room$ & 12.8 & 4.5 \\
\hline
degree-degree correlations $\Pi(\vec{k},\vec{k}^\prime)\room$ & 11.6 & 4.1\\
\hline
\end{tabular}
\end{tabular}
\vsp

\caption{The tailoring of random graph ensembles by imposing as constraints the values of increasingly prescriptive macroscopic 
topological features measured in the gene regulation network of \cite{citeulike:466068}. The tailoring reduces the entropy per node $S$ in the ensemble
in stages, and thereby the effective number of graphs 
${\cal N}=\exp[NS]$ compatible with the network of \cite{citeulike:466068}. 
As in the previous example, refining the tailoring of the graph ensemble 
from imposing only the correct average degree 
to imposing the correct degree distribution is  more significant than the further refinement of imposing the correct degree-degree correlations. 
Hence the degree complexity of this network is again significantly larger than the wiring complexity. }
\label{tab:harbison}
\end{table}

In practice,  when the gene network data are collected, a decision has to be made about the cut-off point where the effect of one gene product on another gene is so small as to be considered insignificant. If there was no threshold and every small fluctuation was taken to be evidence of co-regulation, then it would appear that every gene regulated every other gene, and the network would be complete. Conversely, setting too strict a threshold will risk missing out on important but subtle interactions. 

\begin{figure}[h]
\centerline
{
	\mbox{
			\includegraphics[scale=0.4]{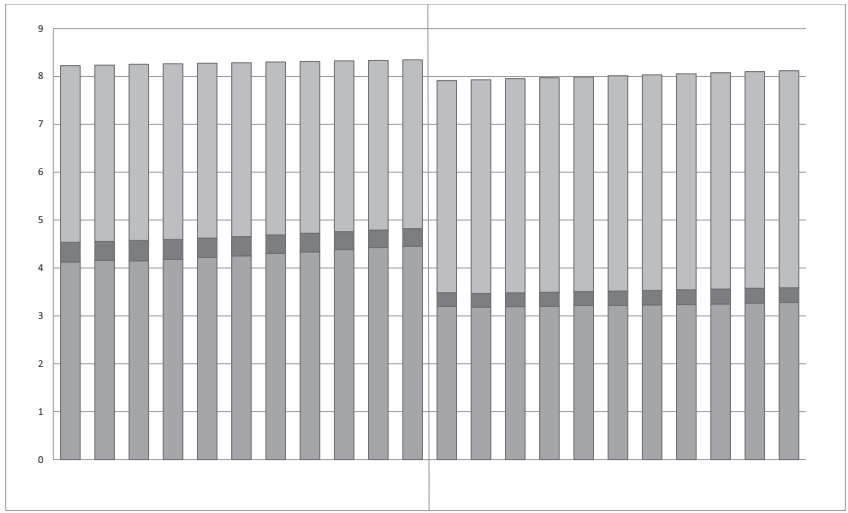}
		}
}
\caption{
Each bar on the chart represents a different choice of threshold. Moving from left to right, the threshold is made progressively stricter so as to exclude approximately 3 percent of arcs at each step. The left half refers to Harbison \textit{et al.} \cite{citeulike:466068} data; the right half refers to Hughes \textit{et al.} \cite{citeulike:124} data. 
Within a bar, the top line presents the entropy per bond when the constraint is `average degree'; the next line shows the entropy per bond when the constraint is additionally `degree distribution'; and, the final line gives the entropy per bond for the ensemble additionally targeting the `degree-degree correlation'. Hence the top two shaded areas represent the degree complexity and the wiring complexity respectively. 
Both datasets are plotted on the same axis in order to illustrate that, although there is some movement with different thresholds, the results for the two different networks remain distinct and distinguishable for any reasonable choice of threshold, and are not unduly sensitive to any reasonable choice of threshold. 
}
\label{fig:thresholds}
\end{figure}

Changing the threshold would reduce the number of arcs, and hence make the network more sparse with lower average degree. Our base assumption would be that beyond that, the main qualitative features of the topology would be maintained. That is, the stricter threshold would remove arcs indiscriminately across the network. However, it is possible that, for example, a node would appear to be a `hub' under a lenient criterion, but would lose a large number of interactions under stricter criteria, so that it is no longer a hub: this would be a qualitative change to the topology arising from the change in thresholds. The analysis proposed in this paper is measuring the topological properties of the network (rather than the network itself). We would expect these results to vary insofar as the topological properties varied. Figure \ref{fig:thresholds} shows the results of repeating the analysis above for different values of the thresholds. 

The above data all refer to the same organism, yeast; however, they present different aspects of gene interactions. Hence, even more than for protein-protein interaction networks, comparison must be done cautiously.  The heterogeneity in the data sets emphasises the importance of developing a suite of tools and measures that can be used to study each network independently.

\section{Discussion}

In this paper we have derived several mathematical results for directed random graph ensembles tailored to match chosen properties of real-world networks. We have calculated the Shannon entropy of ensembles constrained by a prescribed degree distribution, and of ensembles constrained by a prescribed degree-degree correlation function (which contains more detailed topological information than the degree distribution). We have also defined a rational information-theoretic distance measurement for comparing networks based on their degree distribution and degree-degree correlation. 
All this complements and generalises earlier work done in \cite{AnnCooFerFraKle09} for nondirected networks. We also identified 
a correction term to the distance measure of nondirected graphs which was absent in \cite{AnnCooFerFraKle09}.  
A summary of our results and how they compare with the corresponding formula for nondirected networks is presented in tables \ref{tab:compare} and \ref{tab:compare_dist}.

Our growing suite of quantitative tools can be used to study the properties of large real world networks. These tools are precise in leading order in $N$, and take the form of explicit and transparent formulae which use easily measurable macroscopic parameters as input. The present generalization to nondirected networks enables their application to gene regulation networks. We trust that the benefits of having explicit formulae for network complexities and information-theoretic dissimilarity measures will increase, especially in bioinformatics, as we gain experience with using and interpreting the method, and as we increase the range of topological properties to which we can tailor our graph ensembles. 

The focus of our future work will be to increase the number of topological properties that we can
characterise, measure, and impose upon tailored random graph ensembles. Significant progress has already been
made towards including distributions of so-called generalised degrees, but our priority will be to focus on observables that measure the statistics of short loops.
In the presence of such loops the methods and ideas that we applied so far  will no longer suffice. However,
short loops appear to be key biological motifs, so progress in this direction should yield substantial benefits
in terms of applicability of the method in biological signalling. 

\section*{References}

\bibliography{RobertsCoolenSchlitt}{}
\bibliographystyle{visual_match_to_jphysA}

\clearpage
\appendix

\section{Order parameter representation of the graph probabilities}
\label{app:OmegaCalculation}

In this section we derive a tool that is repeatedly used in this paper, being a formula in terms of 
simple observables and order parameters of the log-probability per node of graphs (\ref{eq:Omega}) 
given the ensemble definition (\ref{eq:fullfamily}), in leading orders in $N$. 
Upon substituting  (\ref{eq:fullfamily})  into this formula, and after some simple manipulations and use of the law of large numbers, one finds 
\begin{eqnarray}
\Omega(\bc|p,Q)&=&	 \sum_{\vec{k}}p(\vec{k}|\bc) \log p(\vec{k})+\phi_1(\bc|Q)-\phi_2(\bc|Q) +   \epsilon_N
\label{eq:OmegaApp}
\\
\phi_1(\bc|Q)&=& 
 \frac{1}{N}\log w(\bc|\vec{k}_1,\ldots,\vec{k}_N,Q)\Big|_{\vec{k}_i=\vec{k}_i(\bc)~\forall i}
 \label{eq:phi1}
 \\
 \phi_2(\bc|Q)&=&  
\frac{1}{N}\log  Z(\vec{k}_1,\ldots,\vec{k}_N,Q)\Big|_{\vec{k}_i=\vec{k}_i(\bc)~\forall i}
 \label{eq:phi2}
\end{eqnarray}
with $ \epsilon_N \rightarrow 0$ as $ N\rightarrow \infty $, and 
\begin{eqnarray}
\hspace*{-15mm}
Z(\vec{k}_1,\ldots,\vec{k}_N,Q)&=&  \sum_{\bc}w(\bc|\vec{k}_1,\ldots,\vec{k}_N, Q) \prod_i \delta_{\vec{k}_i,\vec{k}_i(\bc)} 
\\
\hspace*{-15mm}
w(\bc\vert \vec{k}_1,\ldots,\vec{k}_N, Q)&=&
\prod_{i\neq j}\Big[\frac{\overline{k}}{N}Q(\bki,\bkj|\bar{p})\delta_{c_{ij},1}\plus 
\Big(1\minus \frac{\overline{k}}{N}Q(\bki,\bkj|\bar{p})\Big)\delta_{c_{ij},0}\Big]
\end{eqnarray}
In these expressions $\overline{k}=N^{-1}\sum_i k_i^{\rm in}=N^{-1}\sum_i k_i^{\rm out}$,  $\bar{p}(\vec{k})=N^{-1}\sum_i\delta_{\vec{k},\vec{k}_i}$, and the kernel $Q(.,.)$ is normalized locally according to $\sum_{\vec{k},\vec{k}^\prime}\bar{p}(\vec{k})\bar{p}(\vec{k^\prime})Q(\vec{k},\vec{k}^\prime|\bar{p})=1$.

\subsection{Calculation of $\phi_1$}

The first contribution (\ref{eq:phi1}) to the entropy is calculated easily:
\begin{eqnarray}
\hspace*{-15mm}
\phi_1(\bc|Q)&=& 
 \frac{1}{N}\sum_{i\neq j}\Big\{ 
  c_{ij}\log\Big[
  \frac{\overline{k}}{N}Q(\bki,\bkj|\bar{p})\Big]
- \frac{\overline{k}}{N}Q(\bki,\bkj|\bar{p})
\Big\}\Big|_{\vec{k}_i=\vec{k}_i(\bc)~\forall i}
 +\order(\frac{1}{N})~
\nonumber
\\
\hspace*{-15mm}
&=& 
\bar{k}(\bc)\Big\{ \!\log\Big[\frac{\overline{k}(\bc)}{N}\Big]
\!-\!1\!
+\!
\sum_{\vec{k},\vec{k}^\prime}\!W(\vec{k},\vec{k}^\prime|\bc) \log Q(\vec{k},\vec{k}^\prime|p(.|\bc)) \Big\}\!
 +\order(\frac{1}{N})
 \label{eq:final_phi1}
 \end{eqnarray}
 It involves the in- and out degree distribution $p(\vec{k}|\bc)$,  its degree average $\bar{k}(\bc)$, and the joint distribution $W(\vec{k},\vec{k}^\prime|\bc)$ of in- and out degrees of connected nodes. All are calculated for the graph $\bc$ and defined as
\begin{eqnarray}
p(\vec{k}|\bc)&=& \frac{1}{N}\!\sum_i \delta_{\vec{k},\vec{k}(\bc)}\\
W(\vec{k},\vec{k}^\prime|\bc)&=& \frac{1}{N\bar{k}(\bc)}\sum_{ij}c_{ij}\delta_{\vec{k},\vec{k}_i(\bc)}\delta_{\vec{k}^\prime,\vec{k}_j(\bc)}
\label{eq:defineW}
\end{eqnarray}
They are related via the two identities
\begin{eqnarray} 
W_1(\vec{k}|\bc)=\sum_{\vec{k}^\prime}W(\vec{k},\vec{k}^\prime|\bc)=\frac{k^{\rm in}}{\bar{k}(\bc)}p(\vec{k}|\bc)
\label{eq:Wmarginal1}
\\
W_2(\vec{k}|\bc)=\sum_{\vec{k}^\prime}W(\vec{k}^\prime,\vec{k}|\bc)=\frac{k^{\rm out}}{\bar{k}(\bc)}p(\vec{k}|\bc)
\label{eq:Wmarginal2}
\end{eqnarray}
The kernel in  (\ref{eq:final_phi1})
is normalized according to $\sum_{\vec{k},\vec{k}^\prime}p(\vec{k}|\bc)p(\vec{k}^\prime|\bc)Q(\vec{k},\vec{k}^\prime|p(.|\bc))=1$. 

\subsection{Calculation of $\phi_2$}

In order to calculate (\ref{eq:phi2}) we first work out the following quantity, which will then have to be evaluated at 
$(\vec{k}_1,\ldots,\vec{k}_N)=(\vec{k}_1(\bc),\ldots,\vec{k}_N(\bc))$:
\begin{eqnarray}
&&\hspace*{-20mm}
\tilde{\phi}_2(\vec{k}_1,\ldots,\vec{k}_N|Q)=  
\frac{1}{N}\log  Z(\vec{k}_1,\ldots,\vec{k}_N,Q)
\nonumber
\\
&=&  
\frac{1}{N}\log  \sum_{\bc} 
\prod_{i\neq j}\Big[\frac{\overline{k}}{N}Q(\bki,\bkj|\bar{p})\delta_{c_{ij},1}\plus 
\Big(1\minus \frac{\overline{k}}{N}Q(\bki,\bkj|\bar{p})\Big)\delta_{c_{ij},0}\Big] 
\nonumber
\\[-2mm]
&&\hspace*{50mm}\times\prod_i \delta_{\vec{k}_i,\vec{k}_i(\bc)}\nonumber
\\
&=& 
\frac{1}{N}\log 	\int_{\minus \pi }^{\pi } \!\prod_{i} \Big[\frac{\rmd\omega_i\rmd\psi_i}{4\pi^2}\rme^{\rmi[\omega_i k_i^{\rm in}+\psi_i k_i^{\rm out}]}\Big]
L(\bomega, \bpsi|\bar{p},Q)
\end{eqnarray}
with 
\begin{eqnarray}
L(\bomega,\bpsi|\bar{p},Q) &=&
\prod_{i\neq j}\Big[1\!+\!\frac{\overline{k}}{N}Q(\bki,\bkj|\bar{p})[\rme^{-\rmi(\omega_i+\psi_j)}\minus1]\Big] 
\nonumber\\
&=& \exp\Big[\!\frac{\overline{k}}{N}\sum_{ij}Q(\bki,\bkj|\bar{p})[\rme^{-\rmi(\omega_i+\psi_j)}\minus1]\!+\!  \mathcal{O}(N^{0})\Big] 	~			
\end{eqnarray}
Upon introducing $R(\vec{k}|\bomega)=N^{-1}\sum_i \delta_{\vec{k},\vec{k}_i}\rme^{-\rmi\omega_i}$ and 
$S(\vec{k}|\bpsi)=N^{-1}\sum_i \delta_{\vec{k},\vec{k}_i}\rme^{-\rmi\psi_i}$, 
and inserting $\int\!\prod_{\vec{k}}\!\big[\rmd R(\vec{k})\rmd S(\vec{k})~\delta[R(\vec{k})\!-\!R(\vec{k}|\bomega)]\delta[S(\vec{k})\!-\!S(\vec{k}|\bpsi)]\big]$ with $\delta$-functions written in integral form, we can write 
\begin{eqnarray}
\hspace*{-15mm}
L(\bomega,\bpsi|\bar{p},Q) &=& \int\!\prod_{\vec{k}}\Big[\frac{\rmd R(\vec{k})\rmd\hat{R}(\vec{k})\rmd S(\vec{k})\rmd\hat{S}(\vec{k})}{4\pi^2/N^2}
\rme^{\rmi N[\hat{R}(\vec{k})R(\vec{k})+\hat{S}(\vec{k})S(\vec{k})]}
\Big]\rme^{\mathcal{O}(N^{0})}
\nonumber
\\
\hspace*{-15mm}
&&\hspace*{-5mm}\times
\rme^{-\rmi \sum_i [\hat{R}(\vec{k}_i)\rme^{-\rmi\omega_i}+\hat{S}(\vec{k}_i)\rme^{-\rmi\psi_i}]+
 \overline{k}N\sum_{\vec{k},\vec{k}^\prime}R(\vec{k}) Q(\vec{k},\vec{k}^\prime|\bar{p})S(\vec{k}^\prime)
 -\overline{k}N
}						
\end{eqnarray}
Substituting this back into $\tilde{\phi}_2$, and using the law of large numbers, then  gives
\begin{eqnarray}
\hspace*{-15mm}
\tilde{\phi}_2(\ldots)&=&  \frac{1}{N}\!\log\!
 \int\!\prod_{\vec{k}}\Big[\rmd R(\vec{k})\rmd\hat{R}(\vec{k})\rmd S(\vec{k})\rmd\hat{S}(\vec{k})\Big]
\rme^{N\Psi[R,\hat{R},S,\hat{S}|\bar{p},Q]+\mathcal{O}(\log N)}
\end{eqnarray}
where
\begin{eqnarray}
\hspace*{-15mm}
\Psi[R,\hat{R},S,\hat{S}|\bar{p},Q]&=& 
\rmi \sum_{\vec{k}}[\hat{R}(\vec{k})R(\vec{k})\!+\!\hat{S}(\vec{k})S(\vec{k})]
+
\overline{k}\sum_{\vec{k},\vec{k}^\prime}R(\vec{k})Q(\vec{k},\vec{k}^\prime|\bar{p})S(\vec{k}^\prime)
-
 \overline{k}
 \nonumber
 \\
 \hspace*{-15mm}
 &&\hspace*{-16mm}
 +\sum_{\vec{k}}\bar{p}(\vec{k})
  \Big\{
 \log\! \int_{\minus \pi }^{\pi }\!\frac{\rmd\omega}{2\pi}\rme^{\rmi[\omega k^{\rm in}-\hat{R}(\vec{k})\rme^{-\rmi\omega}]
 }
  +\log\! 
 \int_{\minus \pi }^{\pi }\!\frac{\rmd\psi}{2\pi}\rme^{\rmi[\psi k^{\rm out}-\hat{S}(\vec{k})\rme^{-\rmi\psi}]}
 \Big\}
 \nonumber
 \\[-2mm]
 \hspace*{-15mm}&&
 \label{eq:firstPsi}
 \end{eqnarray}
After doing the remaining integrals over $\omega$ and $\psi$ we get
 \begin{eqnarray}
\hspace*{-15mm}
\Psi[R,\hat{R},S,\hat{S}|\bar{p},Q]&=& 
\rmi \sum_{\vec{k}}[\hat{R}(\vec{k})R(\vec{k})\!+\!\hat{S}(\vec{k})S(\vec{k})]
+
\overline{k}\sum_{\vec{k},\vec{k}^\prime}R(\vec{k})Q(\vec{k},\vec{k}^\prime|\bar{p})S(\vec{k}^\prime)
-\overline{k}
 \nonumber
 \\
 \hspace*{-15mm}
 &&
 +\sum_{\vec{k}}\bar{p}(\vec{k})k^{\rm in}
 \log [-\rmi \hat{R}(\vec{k})]
+\sum_{\vec{k}}\bar{p}(\vec{k})k^{\rm out}\log 
  [-\rmi \hat{S}(\vec{k})]
  \nonumber
  \\
  \hspace*{-10mm}&&\hspace*{20mm}
 -\sum_{\vec{k}}\bar{p}(\vec{k})
 \log (k^{\rm in}!k^{\rm out}!)
 \end{eqnarray}
For $N\to\infty$ the quantity $\tilde{\phi}_2(\vec{k}_1,\ldots,\vec{k}_N|Q)$ can be evaluated by steepest descent, giving 
$\lim_{N\rightarrow \infty} 	\tilde{\phi}_2(\ldots)=\extr_{R, \hat{R}, S, \hat{S}} \Psi[R, \hat{R}, S, \hat{S}|\bar{p},Q]$. 
Differentiation of $\Psi$ gives the following saddle-point equations:
 \begin{eqnarray}
-\rmi\hat{R}(\vec{k})
&=& \bar{p}(\vec{k})k^{\rm in}/ R(\vec{k})
=
\overline{k}\sum_{\vec{k}^\prime}Q(\vec{k},\vec{k}^\prime|\bar{p})S(\vec{k}^\prime)
  \\
-\rmi \hat{S}(\vec{k})
&=& \bar{p}(\vec{k})k^{\rm out}/S(\vec{k}) = 
\overline{k}\sum_{\vec{k}^\prime}Q(\vec{k}^\prime,\vec{k}|\bar{p})R(\vec{k}^\prime) 
\end{eqnarray}
At the saddle-point we deduce that $\sum_{\vec{k},\vec{k}^\prime}R(\vec{k})Q(\vec{k},\vec{k}^\prime|\bar{p})S(\vec{k}^\prime)=1$, 
and that 
 \begin{eqnarray}
\hspace*{-10mm}
\Psi[R,\hat{R},S,\hat{S}|\bar{p},Q]&=& 
-
2\bar{k}
 -\sum_{\vec{k}}\bar{p}(\vec{k})
 \log (k^{\rm in}!k^{\rm out}!)
 \nonumber
 \\
 \hspace*{-10mm}
 &&
 \hspace*{-20mm}
 +\sum_{\vec{k}}\bar{p}(\vec{k})k^{\rm in}
 \log \Big[ \frac{\bar{p}(\vec{k})k^{\rm in}}{R(\vec{k}|\bar{p},Q)}\Big]
+\sum_{\vec{k}}\bar{p}(\vec{k})k^{\rm out}\log 
  \Big[\frac{\bar{p}(\vec{k})k^{\rm out}}{S(\vec{k}|\bar{p},Q)}\Big]
  \label{eq:Psi_phi2}
 \end{eqnarray}
in which the functions $R(\vec{k}|\bar{p},Q)$ and $S(\vec{k}|\bar{p},Q)$ are the solutions of
 \begin{eqnarray}
 \hspace*{-10mm}
 R(\vec{k})=\frac{ \bar{p}(\vec{k})k^{\rm in}}
{\overline{k}\sum_{\vec{k}^\prime}Q(\vec{k},\vec{k}^\prime|\bar{p})S(\vec{k}^\prime)},
  ~~~~~~~~
S(\vec{k})= \frac{\bar{p}(\vec{k})k^{\rm out}}
{ \overline{k}\sum_{\vec{k}^\prime}Q(\vec{k}^\prime,\vec{k}|\bar{p})R(\vec{k}^\prime)}
\end{eqnarray}
Finally, the quantity (\ref{eq:phi2}) we aim to calculate 
is defined as the value of $\tilde{\phi}_2(\ldots)$ upon substituting $(\vec{k}_1,\ldots,\vec{k}^N)\to 
(\vec{k}_1(\bc),\ldots,\vec{k}_N(\bc))$. The only occurrences of the sequence $(\vec{k}_1,\ldots,\vec{k}_N)$ in the formula 
  (\ref{eq:Psi_phi2}) are in the values of $\bar{p}(\vec{k})$ and $\bar{k}$, so we obtain  $\phi_2(\bc|Q)$ by making in  (\ref{eq:Psi_phi2}) 
  the substitutions $\bar{p}(\vec{k})\to p(\vec{k}|\bc)$ and $\bar{k}\to \bar{k}(\bc)$. 
We conclude that 
\begin{eqnarray}
\hspace*{-10mm}
 \phi_2(\bc|Q)&=&  
 -
2\tilde{k}
 -\sum_{\vec{k}}\tilde{p}(\vec{k})
 \log (k^{\rm in}!k^{\rm out}!)
 \nonumber
 \\
 \hspace*{-10mm}
 &&\hspace*{-5mm}
 +\sum_{\vec{k}}\tilde{p}(\vec{k})k^{\rm in}
 \log \Big[ \frac{\tilde{p}(\vec{k})k^{\rm in}}{R(\vec{k}|\tilde{p},Q)}\Big]
+\sum_{\vec{k}}\tilde{p}(\vec{k})k^{\rm out}\log 
  \Big[\frac{\tilde{p}(\vec{k})k^{\rm out}}{S(\vec{k}|\tilde{p},Q)}\Big]
  \label{eq:final_phi2}
 \end{eqnarray}
  in which  $\tilde{p}(\vec{k})=p(\vec{k}|\bc)$ and $\tilde{k}=\bar{k}(\bc)$, and 
 in which $R(\vec{k}|\tilde{p},Q)$ and $S(\vec{k}|\tilde{p},Q)$   are the solutions of
\begin{eqnarray}
\hspace*{-10mm}
 R(\vec{k})=\frac{ \tilde{p}(\vec{k})k^{\rm in}}
{\tilde{k}\sum_{\vec{k}^\prime}Q(\vec{k},\vec{k}^\prime|\tilde{p})S(\vec{k}^\prime)},
  ~~~~~~
S(\vec{k})= \frac{\tilde{p}(\vec{k})k^{\rm out}}
{ \tilde{k}\sum_{\vec{k}^\prime}Q(\vec{k}^\prime,\vec{k}|\tilde{p})R(\vec{k}^\prime)}
\end{eqnarray}

\subsection{Final analytical expression for $\Omega$}

The intermediate results (\ref{eq:final_phi1},\ref{eq:final_phi2}) can now be substituted back into 
expression (\ref{eq:OmegaApp}), which gives a formula that is seen to depend on $\bc$ only via 
$W(\vec{k},\vec{k}^\prime|\bc)$ and $p(\vec{k}|\bc)$:
\begin{eqnarray}
\hspace*{-10mm}
\Omega(\bc|p,Q)&=&	\Big\{ 
\sum_{\vec{k}}\tilde{p}(\vec{k}) \log p(\vec{k})+
\tilde{k}\big[1\!+\!\log[\tilde{k}/N]\big]
 +\sum_{\vec{k}}\tilde{p}(\vec{k})
 \log (k^{\rm in}!k^{\rm out}!)
 \nonumber
 \\
 \hspace*{-10mm}
 &&\hspace*{-0mm}
~~ 
 -\sum_{\vec{k}}\tilde{p}(\vec{k})k^{\rm in}
 \log \Big[ \frac{\tilde{p}(\vec{k})k^{\rm in}}{R(\vec{k}|\tilde{p},Q)}\Big]
-\sum_{\vec{k}}\tilde{p}(\vec{k})k^{\rm out}\log 
  \Big[\frac{\tilde{p}(\vec{k})k^{\rm out}}{S(\vec{k}|\tilde{p},Q)}\Big]
  \nonumber
\\[1mm]
\hspace*{-10mm}
&&
\hspace*{0mm}
~~
+~
\tilde{k}\sum_{\vec{k},\vec{k}^\prime}\tilde{W}(\vec{k},\vec{k}^\prime) \log Q(\vec{k},\vec{k}^\prime|\tilde{p}) 
~\Big\}_{\tilde{W}(\vec{k},\vec{k^\prime})=W(\vec{k},\vec{k}^\prime|\bc),~\tilde{p}(\vec{k})=p(\vec{k}|\bc)}
\nonumber
\\
\hspace*{-10mm}&&
 \hspace*{70mm} +   \varepsilon_N
 \label{eq:Omega_result}
\end{eqnarray}
with $\lim_{N\to\infty}\varepsilon_N=0$, $\tilde{k}=\sum_{\vec{k}}k^{\rm in} \tilde{p}(\vec{k})=\sum_{\vec{k}}k^{\rm out} \tilde{p}(\vec{k})$, and with the two functions $S(\vec{k}|\tilde{p},Q)$ and $R(\vec{k}|\tilde{p},Q)$ to be extracted from (\ref{eq:SQeqns}).

\section{Calculation of the kernel $W$}
\label{app:findW}

For large $N$ the kernel $W(\vec{k},\vec{k}^\prime)=(N\bar{k})^{-1}\sum_{ij}c_{ij}\delta_{\vec{k},\vec{k}_i}
\delta_{\vec{k}^\prime,\vec{k}_j}$ will be self-averaging in  the ensemble (\ref{eq:fullfamily}), i.e. with probability one any graph generated randomly according to  (\ref{eq:fullfamily}) will exhibit the same kernel, modulo finite size effects. 
Thus we may for $N\to \infty$ calculate $W(\vec{k},\vec{k}^\prime)$ as an average over the ensemble (\ref{eq:fullfamily}):
\begin{eqnarray}
\hspace*{-15mm}
W(\vec{k},\vec{k}^\prime)&=& \frac{1}{N\bar{k}}\sum_{r\neq s}
 \sum_{\vec{k}_1\ldots \vec{k}_N}
	 \frac{\delta_{\vec{k},\vec{k}_r}\delta_{\vec{k}^\prime,\vec{k}_s}\prod_i p(\vec{k}_i)}{Z(\vec{k}_1\ldots\vec{k}_N,Q)}\sum_{\bc} \Big[\prod_i \delta_{\vec{k}_i,\vec{k}_i(\bc)}\Big]
c_{rs}
	 \nonumber
	 \\
	 \hspace*{-15mm}
	 &&\hspace*{10mm}
	 \times
\prod_{i\neq j}\Big[\frac{\overline{k}}{N}Q(\bki,\bkj|p)\delta_{c_{ij},1}\plus 
\Big(1\minus \frac{\overline{k}}{N}Q(\bki,\bkj|p)\Big)\delta_{c_{ij},0}\Big] 
\nonumber
\\
\hspace*{-15mm}
&=& \frac{1}{N^2}\sum_{r\neq s}
 \sum_{\vec{k}_1\ldots \vec{k}_N}
	 \frac{\delta_{\vec{k},\vec{k}_r}\delta_{\vec{k}^\prime,\vec{k}_s}\prod_i p(\vec{k}_i)}{Z(\vec{k}_1\ldots\vec{k}_N,Q)}
	 	\int_{\minus \pi }^{\pi } \!\prod_{i} \Big[\frac{\rmd\omega_i\rmd\psi_i}{4\pi^2}\rme^{\rmi[\omega_i k_i^{\rm in}+\psi_i k_i^{\rm out}]}\Big]
	 \nonumber
	 \\
	 \hspace*{-15mm}
	 &&
	 \times
Q(\vec{k}_r,\vec{k}_s|p)[\rme^{-\rmi(\omega_r+\psi_s)}[1\!+\!\order(\frac{1}{N})]
\nonumber
\\
&&\times
\prod_{i\neq j}\Big[1\!+\!\frac{\overline{k}}{N}Q(\bki,\bkj|p)[\rme^{-\rmi(\omega_i+\psi_j)}\minus1]\Big] 	
\nonumber
\\
\hspace*{-15mm}
&=&Q(\vec{k},\vec{k}^\prime|p)\!\!
 \sum_{\vec{k}_1\ldots \vec{k}_N}\!\!
	 \frac{\prod_i p(\vec{k}_i)}{Z(\vec{k}_1\ldots\vec{k}_N,Q)}
	 \int_{\minus \pi }^{\pi } \!\prod_{i} \Big[\frac{\rmd\omega_i\rmd\psi_i}{4\pi^2}\rme^{\rmi[\omega_i k_i^{\rm in}+\psi_i k_i^{\rm out}]}\Big]
	 \nonumber
	 \\
	 \hspace*{-15mm}
	 &&
 \times 
L(\bomega,\bpsi|p,Q)	 \Big(\frac{1}{N}\!\sum_{r}\!\delta_{\vec{k},\vec{k}_r}
	 \rme^{-\rmi\omega_r}\Big)\Big( \frac{1}{N}\!\sum_{s}\!\delta_{\vec{k}^\prime,\vec{k}_s}
	 \rme^{-\rmi\psi_s}\Big) [1\!+\!\order(\frac{1}{N})]
\nonumber
\\
\hspace*{-15mm}
&=& Q(\vec{k},\vec{k}^\prime|p)\!\!
 \sum_{\vec{k}_1\ldots \vec{k}_N}\!\!\!
	 \frac{[1\!+\!\order(\frac{1}{N})]\prod_i p(\vec{k}_i)}{Z(\vec{k}_1\ldots\vec{k}_N,Q)}
	\!\! \int\!\prod_{\vec{q}}\Big[\frac{\rmd R(\vec{q})\rmd\hat{R}(\vec{q})\rmd S(\vec{q})\rmd\hat{S}(\vec{q})}{4\pi^2/N^2}
	 \Big]
	 \nonumber
\\
\hspace*{-15mm}&&
\times
\rme^{\rmi N\sum_{\vec{q}}[\hat{R}(\vec{q})R(\vec{q})+\hat{S}(\vec{q})S(\vec{q})]
+ \overline{k}N\sum_{\vec{q},\vec{q}^\prime}Q(\vec{q},\vec{q}^\prime|p)R(\vec{q})S(\vec{q}^\prime)
 -  \overline{k}N
+\mathcal{O}(N^{0})}
	 \nonumber
	 \\
	 \hspace*{-15mm}
	 &&
 \times R(\vec{k})S(\vec{k}^\prime) \prod_i\int_{\minus \pi }^{\pi } \! \Big[\frac{\rmd\omega\rmd\psi}{4\pi^2}\rme^{\rmi\omega k_i^{\rm in}+\rmi\psi k_i^{\rm out}
 -\rmi\hat{R}(\vec{k}_i)\rme^{-\rmi\omega}-\rmi\hat{S}(\vec{k}_i)\rme^{-\rmi\psi}}\Big]			
\end{eqnarray}
We now write $Z(\vec{k}_1\ldots\vec{k}_N,Q)$ also as an integral over order parameters, as in our earlier derivation of  (\ref{eq:Psi_phi2}), 
but noting that now the relevant degree distribution is that of our ensemble (\ref{eq:fullfamily}), i.e. $p(\vec{k})$ instead of $\bar{p}(\vec{k})$. 
This gives 
\begin{eqnarray}
\hspace*{-10mm}
W(\vec{k},\vec{k}^\prime) &=& [1\!+\!\order(\frac{1}{N})]Q(\vec{k},\vec{k}^\prime)
 \sum_{\vec{k}_1\ldots \vec{k}_N}\!\!\prod_i p(\vec{k}_i)
	 \nonumber
\\
\hspace*{-10mm}
&&\hspace*{-15mm}
\times
\frac{
 \int\!\prod_{\vec{q}}\rmd R(\vec{q})\rmd\hat{R}(\vec{q})\rmd S(\vec{q})\rmd\hat{S}(\vec{q})
~\rme^{N\Psi[R,\hat{R},S,\hat{S}|p,Q]+\mathcal{O}(\log N)}
R(\vec{k})S(\vec{k^\prime)}}
{
 \int\!\prod_{\vec{q}}\rmd R(\vec{q})\rmd\hat{R}(\vec{q})\rmd S(\vec{q})\rmd\hat{S}(\vec{q})
~\rme^{N\Psi[R,\hat{R},S,\hat{S}|p,Q]+\mathcal{O}(\log N)}}		
\end{eqnarray}
where the non-extensive terms in the exponentials of numerator and denominator are fully identical, and with $\Psi$ as defined 
in (\ref{eq:firstPsi}), modulo the replacement $\bar{p}\to p$. The summation over degree sequences has now become obsolete, and for $N\to\infty$ we obtain
\begin{eqnarray}
\lim_{N\to\infty}
W(\vec{k},\vec{k}^\prime) &=& R(\vec{k}|p,Q) Q(\vec{k},\vec{k}^\prime|p)
	S(\vec{k}^\prime|p,Q)
\label{eq:Wfound}	
\end{eqnarray}
in which $R(\vec{k}|p,Q)$ and $S(\vec{k}|p,Q)$ are to be solved from 
 \begin{eqnarray}
 \hspace*{-5mm}
 R(\vec{k})=\frac{ p(\vec{k})k^{\rm in}}
{\overline{k}\sum_{\vec{k}^\prime}Q(\vec{k},\vec{k}^\prime|p)S(\vec{k}^\prime)},
  ~~~~~~~~
S(\vec{k})= \frac{p(\vec{k})k^{\rm out}}
{ \overline{k}\sum_{\vec{k}^\prime}Q(\vec{k}^\prime,\vec{k}|p)R(\vec{k}^\prime)}
\end{eqnarray}
with the average degree of our ensemble, $\bar{k}=\sum_{\vec{k}}k^{\rm in}p(\vec{k})=\sum_{\vec{k}}k^{\rm out}p(\vec{k})$. 

\end{document}